\documentclass[12pt,letterpaper,a4paper]{article}
\usepackage[includeheadfoot,
            marginratio={1:1,2:3},
            width=450pt,
            height=700pt,]{geometry}
\usepackage{amsmath}
\usepackage{amsfonts}
\usepackage{amssymb}
\usepackage{graphicx}
\usepackage{cancel}

\usepackage{empheq}
\usepackage{color}
\usepackage{hyperref}

\usepackage{float}
\restylefloat{table}
\restylefloat{figure}

\newcommand{\nc}{\newcommand}
\nc{\beq}{\begin{equation}}
\nc{\eeq}{\end{equation}}
\nc{\bea}{\begin{eqnarray}}
\nc{\eea}{\end{eqnarray}}

\def\ov{\overline}

\allowdisplaybreaks[2]
\numberwithin{equation}{section}

\begin{document}

\vspace{1.5cm}
\begin{center}
{\LARGE
Implementing odd-axions in dimensional oxidation of  4D non-geometric type IIB scalar potential}
\vspace{0.4cm}
\end{center}

\vspace{0.35cm}
\begin{center}
Pramod Shukla \footnote{Email: pkshukla@to.infn.it}
\end{center}

\vspace{0.1cm}
\begin{center}
{
Universit\'a di Torino, Dipartimento di Fisica and I.N.F.N.-sezione di Torino
\vskip0.02cm
via P. Giuria 1, I-10125 Torino, Italy
}
\end{center}

\vspace{1cm}


\begin{abstract}
In a setup of type IIB superstring compactification on an orientifold of a ${\mathbb T}^6/{\mathbb Z}_4$ sixfold, the presence of geometric flux ($\omega$) and non-geometric fluxes ($Q, R$) is implemented along with the standard NS-NS and RR three-form fluxes ($H, F$). After computing the F/D-term contributions to the ${\cal N}=1$ four dimensional effective scalar potential, we rearrange the same into `suitable' pieces by using a set of new generalized flux orbits. Subsequently, we dimensionally oxidize the various pieces of the total four dimensional scalar potential to guess their ten-dimensional origin.
\end{abstract}

\clearpage

\tableofcontents



\section{Introduction}
\label{sec:intro}
String compactifications and gauged supergravities have quite remarkable connections via relating the background fluxes in the former picture with the possible gaugings in the later one  \cite{ Derendinger:2004jn, Derendinger:2005ph, Shelton:2005cf, Samtleben:2008pe, Dall'Agata:2009gv, Aldazabal:2011yz,Dibitetto:2012rk, Villadoro:2005cu, Aldazabal:2006up}. Application of successive $T$-duality operations on three-form $H$-flux of type II orientifold theories results in various geometric and non-geometric fluxes, namely $\omega, \, Q$ and R-fluxes. Moreover, in a setup of type IIB superstring theory compactified on ${\mathbb T}^6/{\left({\mathbb Z}_2 \times {\mathbb Z}_2\right)}$, it was argued that additional fluxes are needed to ensure S-duality invariance of the underlying low energy type IIB supergravity, and in this regard, a new type of non-geometric flux, namely the $P$-flux, has been proposed as a S-dual candidate for the non-geometric $Q$-flux \cite{Aldazabal:2006up,Aldazabal:2008zza,Guarino:2008ik}. The resulting modular completed fluxes can be arranged into spinor representations of $SL(2,{\mathbb Z})^7$, and the compactification manifold with $T$- and $S$-duality appears to be an $U$-fold \cite{Hull:2004in,  Kumar:1996zx, Hull:2003kr} where local patches are glued by performing $T$- and $S$-duality transformations. Since fluxes can induce potentials for various four-dimensional scalars, the same are useful for moduli stabilization and constructing string vacua, and hence connections with gauged supergravity provide a channel to look into phenomenological window, see \cite{Font:2008vd,deCarlos:2009qm, Blumenhagen:2015kja} and references therein. Moreover, in recent years, non-geometric setups have been found to be useful for hunting de-Sitter solutions as well as for building inflationary models \cite{Danielsson:2012by, Blaback:2013ht, Damian:2013dq, Damian:2013dwa, Hassler:2014mla, Blumenhagen:2015qda}.   A consistent incorporation of various kinds of possible fluxes makes the compactification background richer and more flexible for model building. 


Although the origin of all the geometric and non-geometric flux-actions from a ten-dimensional point of view still remains to be (clearly) understood, there have been significant amount of phenomenology oriented studies via considering the 4D effective potential merely derived by knowing the K\"ahler and super-potentials. However, some significant steps have been taken towards exploring the form of non-geometric 10D action via Double Field Theory (DFT) \cite{Andriot:2011uh} \footnote{For recent reviews and more details on flux formulation of DFT, see \cite{Aldazabal:2011nj,Geissbuhler:2011mx,Grana:2012rr}.} as well as supergravity \cite{Villadoro:2005cu, Blumenhagen:2013hva, Gao:2015nra} \footnote{Related to the study of ten-dimensional non-geometric action, see also \cite{Andriot:2013xca,Andriot:2014qla,Sakatani:2014hba} in $\beta$-supergravity framework as well as \cite{Blair:2014zba} for exceptional field theory.}. In this regard, toroidal orientifolds have been always in the center of attraction because of their relatively simpler structure.  Moreover, unlike the case with Calabi Yau compactifications, the explicit and analytic form of metric being known for the toroidal compactification backgrounds make such backgrounds automatically the favorable ones for performing explicit computations. Therefore, the simple toroidal setups have served as promising toolkits for investigating the effects of non-geometric fluxes and also in studying their deeper insights via taking baby steps towards knowing their ten dimensional origin. For example the knowledge of metric has helped in anticipating the ten-dimensional origin of the geometric flux dependent \cite{Villadoro:2005cu} as well as the non-geometric flux dependent potentials \cite{Blumenhagen:2013hva,Gao:2015nra}. Considering a general form of superpotential with the presence of $H$, $\omega$, $Q$, $R$-fluxes in a simple ${\mathbb T}^6/{\left({\mathbb Z}_2 \times {\mathbb Z}_2\right)}$ toroidal orientifold of type IIA and its T-dual type IIB model, the subsequently induced four dimensional scalar potentials have been oxidized into a set of respective pieces of an underlying ten-dimensional supergravity action \cite{Blumenhagen:2013hva}. This {\it dimensional oxidation process} has suggested some peculiar flux combinations to be useful in the ten-dimensional picture, and the same have been further extended with the inclusion of $P$-flux, the S-dual to non-geometric $Q$-flux in \cite{Gao:2015nra}. {\it In addition, with recent attractions triggered in developing axionic models of inflation after BICEP2 and PLANCK \cite{Ade:2014xna, Ade:2015lrj,Ade:2015tva}, a generalization of \cite{Blumenhagen:2013hva, Gao:2015nra} to include involutively odd axions $B_2$ and $C_2$ is desirable not only from the point of view of seeking better understanding of the non-geometric 10D action but also for axionic inflation model building. }For explicit construction of some type-IIB toroidal/CY orientifold examples with odd-axions, see \cite{Lust:2006zg,Lust:2006zh,Blumenhagen:2008zz,Cicoli:2012vw,Cicoli:2012bi,Gao:2013rra,Gao:2013pra}.

Motivated by these aspects, in this article, we implement the presence of odd-axions in the dimensional oxidation process of \cite{Blumenhagen:2013hva} via considering the untwisted sector of type IIB superstring theory compactified on an orientifold of ${\mathbb T}^6/{{\mathbb Z}_4}$. {\it This setup happens to be nontrivial enough as compared to the mostly studied toroidal example of ${\mathbb T}^6/{\left({\mathbb Z}_2 \times {\mathbb Z}_2\right)}$-orientifold in two sense: (i) Having $h^{1,1}_-(X) = 2$, it can accommodate the involutively odd axions, and hence can have the structure of usual flux orbits being corrected via $B_2$-axions similar to type IIA compactification on ${\mathbb T}^6/{\left({\mathbb Z}_2 \times {\mathbb Z}_2\right)}$-orientifold case \cite{Villadoro:2005cu, Blumenhagen:2013hva}; and (ii). it can induce D-terms involving non-geometric R-fluxes also due to non-trivial even (2,1)-cohomology as $h^{2,1}_+(X) = 1$}. On top of these, this setup represents the case of frozen complex structure moduli as $h^{2,1}_-(X) = 0$, and hence is simple enough for explicit computations. With these ingredients, the present toroidal setup provides some interesting and enlightening features of ten-dimensional origin of the 4D non-geometric type IIB scalar potential. 

The paper is organized as follows: the section \ref{sec_Setup} provides some relevant features of type IIB orientifold compactifications followed by an explicit example of  ${\mathbb T}^6/{{\mathbb Z}_4}$-orientifold. In section \ref{sec_potential}, we compute the full scalar potential via considering $F$- and $D$-term contributions. In addition, we invoke the various corrections to flux orbits induced by inclusion of odd axions. Using these new flux-orbits, in section \ref{sec_oxidation}, we first rearrange the total scalar potential into `suitable' pieces which are subsequently oxidized into a ten-dimensional non-geometric action. Finally we conclude in section \ref{sec_conclusion} with a short appendix \ref{sec_components} providing various components of  fluxes/moduli allowed under the orientifold action.

\section{Setup}
\label{sec_Setup}
\subsection{Type-IIB orientifolds and splitting of various cohomologies}
Let us consider Type IIB superstring theory compactified on an orientifold of a
Calabi-Yau threefold $X$.
The admissible orientifold projections fall into two categories,
which are distinguished by their action on the
K\"ahler form $J$ and the holomorphic three-form $\Omega_3$ of
the Calabi-Yau \cite{Grimm:2004uq}:
\begin{eqnarray}
\label{eq:orientifold}
 {\cal O}= \begin{cases}
                       \Omega_p\, \sigma \qquad &{\rm with} \quad
                       \sigma^*(J)=J\,,\quad  \sigma^*(\Omega_3)=\Omega_3 \, ,\\[0.1cm]
                       (-)^{F_L}\,\Omega_p\, \sigma\qquad & {\rm with} \quad
        \sigma^*(J)=J\,, \quad \sigma^*(\Omega_3)=-\Omega_3
\end{cases}
\end{eqnarray}
where $\Omega_p$ is the world-sheet parity transformation and $F_L$ denotes
 the left-moving space-time fermion  number.
Moreover,  $\sigma$ is a holomorphic, isometric
 involution. The first choice leads to orientifold $O9$- and $O5$-planes
whereas the second choice to $O7$- and $O3$-planes.
The $(-)^{F_L}\,\Omega_p\, \sigma$  invariant states in four-dimensions
are listed in table \ref{tableprojection}.
\begin{table}[H]
  \centering
  \begin{tabular}{c||c|c|c|c|c|c}
  & $\phi$&  $g_{\mu \nu}$ &  $B_2$&  $C_0$ & $C_2$ & $C_4$\\
  \hline \hline
 $(-)^{F_L}$ &  $+$  &   $+$  &  $+$   &   $-$    &   $-$  & $-$\\
 $\Omega_p$ &   $+$  &   $+$  &  $-$   &   $-$    &   $+$  & $-$ \\
 $\sigma^*$ &   $+$  &   $+$  &  $-$   &   $+$    &   $-$  & $+$ \\
 \end{tabular}
  \caption{\small Orientifold invariant states.}
  \label{tableprojection}
\end{table}
\noindent
The massless states in the four dimensional effective theory are in one-to-one correspondence
with harmonic forms which are either  even or odd
under the action of $\sigma$. Moreover, these do  generate the equivariant  cohomology groups $H^{p,q}_\pm (X)$. Let us fix the following conventions for the bases of various equivariant cohomologies counting the massless spectra,
\begin{itemize}
 \item {The zero-form: ${\bf 1}$, which is even under $\sigma$.}
 \item {The even two-forms: $\mu_\alpha$, counted by $\alpha=1,...., h^{1,1}_+$.}
 \item {The odd two-forms: $\nu_a$, counted by $a=1,...., h^{1,1}_-$.}
 \item {The even four-forms: $\tilde{\mu}_\alpha$, counted by $\alpha=1,...., h^{1,1}_+$.}
 \item {The odd four-forms: $\tilde{\nu}_a$, counted by $a=1,...., h^{1,1}_-$.}
 \item {A six-form: $\Phi_6 = dx^1 \wedge dx^2 \wedge dx^3 \wedge dx^4 \wedge dx^5 \wedge dx^6$, which is even under $\sigma$.}
\end{itemize}
Here, we take the following definitions of integration over the intersection of various cohomology bases,
\bea
\label{eq:intersection}
& & \int_X \Phi_6 = f, \, \, \int_X \, \mu_\alpha \wedge \tilde{\mu}^\beta = \hat{d}_\alpha^{\, \, \, \beta} , \, \, \int_X \, \nu_a \wedge \tilde{\nu}^b = {d}_a^{\, \, \,b} \nonumber\\
& & \int_X \, \mu_\alpha \wedge \mu_\beta \wedge \mu_\gamma = k_{\alpha \beta \gamma}, \, \, \, \, \int_X \, \mu_\alpha \wedge \nu_a \wedge \nu_b = \hat{k}_{\alpha a b}
\eea
Note that if the four-form basis is chosen to be dual of the two-form basis, one will of course have $\hat{d}_\alpha^{\, \, \, \beta} = \hat{\delta}_\alpha^{\, \, \, \beta}$ and ${d}_a^{\, \, \,b} = {\delta}_a^{\, \, \,b}$. However for the present work, we follow the conventions of \cite{Robbins:2007yv}, and take the generic case. In addition to the splitting of $H^2(X)$ and its dual $H^4(X)$-cohomologies, we also need to know the splitting of three-form cohomology $H^3(X)$ into even/odd eigenspaces under a given involution $\sigma$. Considering the symplectic basis for these even and odd cohomologies $H^3_+(X)$ and $H^3_-(X)$ of three-forms as symplectic pairs $(a_K, b^K)$ and $({\cal A}_k, {\cal B}^k)$ respectively, we fix
\bea
\int_X a_K \wedge b^J = \delta_K{}^J, \, \, \, \, \, \int_X {\cal A}_k \wedge {\cal B}^j = \delta_k{}^j
\eea
Here, for the orientifold choice with $O3/O7$-planes, set of values $\{J,K\}\in \{1, ..., h^{2,1}_+\}$ counting the vector multiplet while $\{j,k\}\in \{0, ..., h^{2,1}_-\}$ counts the number of complex structure moduli. For orientifolds with $O5/O9$-planes, the counting of indices goes as $\{J, K\}\in \{0, ..., h^{2,1}_+\}$ and $\{j,k\}\in \{1, ..., h^{2,1}_-\}$.

Now, the various field ingredients can be expanded in appropriate bases of the equivariant cohomologies. For example, the K\"{a}hler form $J$, the
two-forms $B_2$,  $C_2$ and the R-R four-form $C_4$ can be expanded as \cite{Grimm:2004uq}
\bea
\label{eq:fieldExpansions}
& & J = t^\alpha\, \mu_\alpha,  \,\,\,\quad  B_2= b^a\, \nu_a , \,\,\, \quad C_2 =c^a\, \nu_a \, \\
& & C_4 = D_2^{\alpha}\wedge \mu_\alpha + U^{K}\wedge a_K + U_{K}\wedge b^K + {\rho}_{\alpha} \, \tilde\mu^\alpha \nonumber
\eea
where $t^\alpha$ is two-cycle volume moduli, while $b^a, \, c^a$ and $\rho_\alpha$ are various axions. Further, ($U^K$, $U_K$) forms a dual pair of space-time one forms
and $D_2^{\alpha}$ is a space-time two-form dual to the scalar field $\rho_\alpha$. Due to the self-duality of the R-R four-form, half of
the degrees of freedom of $C_4$ are removed.
Note that the even component of the Kalb-Ramond field
$B_{+} = b^\alpha \, \mu_\alpha$, though  not a continuous modulus,
can take the two  discrete values $b^\alpha\in\{0, 1/2\}$. Further, since $\sigma^*$ reflects the holomorphic three-form,
in the orientifold we have
$h^{2,1}_-(X)$ complex structure moduli $z^{\tilde a}$ appearing as complex scalars. Finally, one has the following table summarizing the ${\cal N}=1$ supersymmetric
massless bosonic spectrum \cite{Grimm:2004uq}, 
\begin{table}[H]
\centering
\begin{tabular}{|l|l|l|}
\hline
 &  & \\[-0.2cm]
  & $h_-^{2,1}$ & $z^{\tilde a}$ \\
chiral multiplets & $h_+^{1,1}$ & $(t^\alpha, \rho_{\alpha})$ \\
  & $h_-^{1,1}$ & $(b^a, c^a)$\\
   & 1 & $(\phi, C_0)$\\
    &  & \\[-0.2cm]
    \hline
    &  & \\[-0.2cm]
vector multiplet & $h_+^{2,1}$ & $U^K$ \\
&  & \\[-0.2cm]
\hline
 &  & \\[-0.2cm]
gravity multiplet & 1 & $g_{\mu\nu}$ \\[0,2cm]
\hline
\end{tabular}
\caption{${\cal N}=1$ massless bosonic spectrum of Type IIB Calabi Yau
  orientifold
\label{table_susy}}
\end{table}
\noindent

Using the pieces of information developed so far, one can collect a complex multi-form of even degree $\Phi_c^{even}$ defined as under \cite{Benmachiche:2006df, Grana:2006hr},
\bea
\label{eq:multiDegreeform}
& & \hskip-2cm \Phi_c^{even} = e^{B_2} \wedge C_{RR} + i \, e^{-\phi} Re(e^{B_2+i\, J})\\
& & \hskip-1cm  =(C_0 + i \, e^{-\phi}) + \left(C_2 + (C_0 + i \, e^{-\phi}) B_2\right) \nonumber\\
& & + \left(C_4^{(0)}+C_2\wedge B_2 + \frac{1}{2} (C_0 + i \, e^{-\phi}) B_2 \wedge B_2 - \frac{i}{2} e^{-\phi} J \wedge J\right)\nonumber \\
& & \hskip-1cm \equiv \tau + G^a \, \,\nu_a + T_\alpha \, \, \tilde{\mu}^\alpha \nonumber
\eea
This suggests the following forms for the appropriate chiral variables appearing as the complex bosons in the respective ${\cal N}=1$ chiral superfields, 
\bea
\label{eq:N=1_coords}
& & \tau = C_0 + \, i \, e^{-\phi} \, \, , \, \, \, \, \, G^a= c^a + \tau \, b^a \, ,\\
& & T_\alpha= \left({\rho}_\alpha +  \, \hat{\kappa}_{\alpha a b} c^a b^b + \frac{1}{2} \, \tau \, \hat{\kappa}_{\alpha a b} b^a \, b^b \right)  -\, \frac{i}{2} \, \kappa_{\alpha\beta\gamma}\, t^\beta t^\gamma   \,.\nonumber
\eea
Here, we have changed the four-cycle volume moduli into Einstein-frame by absorbing $e^{-{\phi}}$ factor (appearing in $\frac{i}{2} e^{-\phi} J \wedge J$) in eqn. (\ref{eq:multiDegreeform}) via redefining the two-cycle volume moduli as $J_E = e^{-{\phi}/2} J$. In the definition of variable $T_\alpha$, we have dropped in index `E' in $t^\alpha$. Also a redefinition of the intersection numbers as compared to the ones given in the definitions of eqn.(\ref{eq:intersection}) is made as: $\kappa_{\alpha\beta\gamma}=(\hat{d^{-1}})_\alpha^{ \, \,\delta} \, k_{\delta\beta\gamma}$ and $\hat{\kappa}_{\alpha a b} = (\hat{d^{-1}})_\alpha^{ \, \,\delta} \, \hat{k}_{\delta a b}$.

The low energy effective action at second order in derivatives is
given  by a supergravity theory, whose dynamics is encoded
in a K\"{a}hler potential $K$, a holomorphic superpotential $W$ and the
holomorphic gauge kinetic functions. These building blocks are written in terms of the aforementioned appropriate chiral variables. 
In our case of present interest, the generic form of K\"{a}hler potential (at tree level) is given as,
\bea
\label{eq:K}
K = - \ln\left(-i(\tau-\ov\tau)\right)
-\ln\left(i\int_{X}\Omega_3\wedge{\bar\Omega_3}\right)-2\ln\left({\cal V}_E\,
(\tau, G^a, T_\alpha; \ov \tau, \ov G^a, \ov T_\alpha)\right)
\eea
where ${\cal V}_E$ is the Einstein frame volume of the Calabi-Yau manifold.
Unfortunately, ${\cal V}_E$ is only implicitly given in terms
of the chiral superfields as it is, in general, non-trivial
to invert the last relation in \eqref{eq:N=1_coords}. 

To express the various geometric as well as non-geometric fluxes into the suitable orientifold even/odd bases, it is important to note that in a given setup, all flux-components will not be generically allowed under the full orientifold action ${\cal O} = \Omega_p (-)^{F_L} \sigma$ \cite{Shelton:2005cf,Aldazabal:2006up}. For example, under the effect of $(\Omega_p (-)^{F_L})$, only geometric flux $\omega$ and non-geometric flux $R$ remain invariant while the standard fluxes $F, H$ and non-geometric $Q$-flux are anti-invariant \cite{Shelton:2005cf,Aldazabal:2006up}. Therefore, under the full orientifold action, we can only have the following components of the standard fluxes $(F, H)$ and the geometric as well as non-geometric fluxes $(\omega, Q$ and $R$),
\bea
& & F\equiv \left(F_k, F^k\right),  H\equiv \left(H_k, H^k\right), \omega\equiv \left({\omega}_a{}^k, {\omega}_{a k} ,\, \hat{\omega}_\alpha{}^K, \hat{\omega}_{\alpha K}\right),\nonumber\\
& & \quad R\equiv \left(R_K, R^K \right), \, \, \, Q\equiv \left({Q}^{a{}K}, \, {Q}^{a}{}_{K}, \, \hat{Q}^{\alpha{}k} , \, \hat{Q}^{\alpha}{}_{k}\right), 
\eea
The structure in which the presence of these flux-components is manifest, can be arranged via the possible three-form components as under \cite{Robbins:2007yv},
\bea
\label{eq:action1}
& & H = {H}^k {\cal A}_k + H_k \, \,{\cal B}^k, \, \, \, \, \, F = {F}^k {\cal A}_k + F_k \, \,{\cal B}^k,\\
& & \hskip-2cm \omega_a \equiv (\omega \triangleleft \nu_a) = {\omega}_a{}^k \, {\cal A}_k + \omega_{a{}k} {\cal B}^k, \, \, \, \, \,\, \, \, \, \hat{Q}^{\alpha}\equiv (Q \triangleright {\tilde\mu}^\alpha) = \hat{Q}^{\alpha{}k} {\cal A}_k + \hat{Q}^{\alpha}{}_{k} {\cal B}^k \nonumber\\
& & \hskip-2cm  \hat{\omega}_\alpha \equiv (\omega \triangleleft \mu_\alpha) = \hat{\omega}_\alpha{}^K a_K + \hat{\omega}_{\alpha{}K} b^K, \, \, \, \, \, {Q}^{a}\equiv (Q \triangleright \tilde{\nu}^a) = {Q}^{a{}K} \, a_K + Q^{a}{}_{K} b^K, \nonumber\\
& & \quad \quad \quad R\bullet \Phi = R^K a_K + R_K b^K\nonumber
\eea
These are relevant for writing down the superpotential contribution as we will see in a moment. Moreover, with these definitions, we have the following non-trivial actions of fluxes on various 3-form even/odd basis elements \cite{Robbins:2007yv},
\bea
\label{eq:action2}
& & H \wedge {\cal A}_k = - f^{-1} H_k \Phi_6, \, \, \quad  \quad \, \quad \quad \, \, H \wedge {\cal B}^k = f^{-1} H^k \Phi_6 \\
& & \omega\triangleleft {\cal A}_k=\left({d}^{-1}\right)_a{}^b \,{\omega}_{b k} \, \tilde{\nu}^a, \, \quad  \quad\, \quad \omega\triangleleft {\cal B}^k=-\left({d}^{-1}\right)_a{}^b \, {\omega}_{b}{}^{k} \, \tilde{\nu}^a\nonumber\\
& & Q\triangleright {\cal A}_k=-\left(\hat{d}^{-1}\right)_\alpha{}^\beta \,\hat{Q}^\alpha_{k} \, {\mu}_\beta, \, \quad  \quad\, Q\triangleright {\cal B}^k=\left(\hat{d}^{-1}\right)_\alpha{}^\beta \, \hat{Q}^{\alpha k} \,{\mu}_\beta ,\nonumber
\eea
and
\bea
\label{eq:action3}
& & R\bullet a_K = f^{-1} \, R_K \, {\bf 1}, \, \, \,\quad  \quad \quad \quad \quad\, \, R \bullet b^K = - f^{-1} \, R^K \,{\bf 1}\\
& & \omega\triangleleft a_K=\left(\hat{d}^{-1}\right)_\alpha{}^\beta \hat{\omega}_{\beta K} \, \tilde{\mu}^\alpha, \, \quad  \quad \, \omega\triangleleft b^K=-\left(\hat{d}^{-1}\right)_\alpha{}^\beta \hat{\omega}_{\beta}{}^{K} \, \tilde{\mu}^\alpha\nonumber\\
& & Q\triangleright a_K=-\left({d}^{-1}\right)_a{}^b {Q}^a_{K} \,{\nu}_b, \,\quad  \quad\, Q\triangleright b^K=\left({d}^{-1}\right)_a{}^b \,{Q}^{a K} \, {\nu}_b . \nonumber
\eea
For writing the flux-superpotential, we further need to define a twisted differential operator, ${\cal D}$ involving the action from all the NS-NS geometric as well as non-geometric fluxes. Following the conventions of \cite{Robbins:2007yv}, the same can be expressed as, 
\bea
\label{eq:twistedD}
& & {\cal D} = d + H \wedge.  - \omega \triangleleft . + Q \triangleright. - R \bullet .
\eea
Now, a generic form of flux superpotential, which includes all the allowed geometric as well as non-geometric flux contributions, can be considered as,
\bea
\label{eq:W1}
& & \hskip-1cm W = \int_X \biggl[F+ {\cal D} \Phi_c^{even}\biggr]_3 \wedge \Omega_3 \\
& & \quad =\int_{X} \biggl[{F} +\tau \, {H} + \, \omega_a \, {G}^a + \, {\hat Q}^{\alpha} \, \,{T}_\alpha \biggr]_3 \wedge \Omega_3. \nonumber
\eea
Note that, only odd-$\omega_a$ and even-$\hat{Q}^\alpha$ components of geometric and non-geometric fluxes are allowed by the choice of involution to contribute into the superpotential. Further, the holomorphic three-form, $\Omega_3$ which is odd under involution, can be generically written in terms of coordinate- and period- vectors in the symplectic basis (${\cal A}_k, {\cal B}^k$) as under,
\bea
\label{eq:Omega3main}
\Omega_3 = {\cal Z}^k {\cal A}_k - {\cal F}_k \, \, {\cal B}^k
\eea
Using the definitions of various flux-actions given in (\ref{eq:action1}), we have the following expansion of the three form appearing in (\ref{eq:W1}),
\bea
& & \left({F} +\tau \, {H} + \, \omega_a \, {G}^a + \, \hat{Q}^{\alpha} \, \,{T}_\alpha \right)\\
& & \quad \quad =\quad \left(F^k + \tau\, H^k + \omega_a{}^k \, G^a + \hat{Q}^{\alpha k} \, T_\alpha\right){\cal A}_k +\left(F_k + \tau\, H_k + \omega_{ak} \, G^a + \hat{Q}^{\alpha}{}_{k} \, T_\alpha\right){\cal B}^k \nonumber
\eea
Subsequently, one arrives at the following generic form of the superpotential
\bea
\label{eq:Wmain}
& & \hskip-1.6cm W = -\biggl[\left(F_k + \tau\, H_k + \omega_{ak} \, G^a + \hat{Q}^{\alpha}{}_{k} \, T_\alpha\right)\, {\cal Z}^k \\
& & \hskip2.5cm + \left(F^k + \tau\, H^k + \omega_a{}^k \, G^a + \hat{Q}^{\alpha k} \, T_\alpha\right) \, {\cal F}_k\biggr]\, . \nonumber
\eea
As also observed in \cite{Robbins:2007yv,Blumenhagen:2015kja}, one should note that R-flux does not appear in the superpotential. In the absence of non-geometric P-flux which is S-dual to Q-fluxes, this form of superpotential is generic at the tree level.

\subsection{An explicit example: Type IIB on a ${\mathbb T}^6/{\mathbb Z}_4$-orientifold}
\label{sec:setup}
We consider the type IIB superstring theory compactified on a toroidal orbifold ${\mathbb T}^6/{\mathbb Z}_4$ with the following redefinition of complexified coordinates on ${\mathbb T}^6$ \cite{Robbins:2007yv},
\begin{eqnarray}
\label{eq:coordinates}
& & z^1 = x^1 + i \, x^2 +e^{i \pi /4}\, (x^3 + i \, x^4) \nonumber\\
& & z^2 = x^3 + i \, x^4 +e^{i 3\, \pi /4}\, (x^1 + i \, x^2) \nonumber\\
& & z^3 = x^5 + i \, x^6 
\end{eqnarray}
The orbifold action ${\mathbb Z}_4$ is given as
\begin{equation}
\Theta({\mathbb Z}_4): (z^1, z^2, z^3) \longrightarrow (i \, z^1, i \, z^2, - z^3)
\end{equation}
The holomorphic involution $\sigma$ is chosen to be,
\begin{eqnarray}
\label{eq:involution}
& & \sigma : (z^1, z^2, z^3) \longrightarrow (- e^{i \, \pi /4}\, z^1, e^{i \, \pi /4}\, z^2, - i \, z^3)
\end{eqnarray}
The hodge number for ${\mathbb T}^6/{\mathbb Z}_4$ orbifold is $h^{2,1}=1$ and $h^{1,1}=5$ which results in splitting into even/odd eigenspaces of (1,1)-cohomology with $h^{1,1}_+=3$ and $h^{1,1}_-=2$ and those of (2,1)-cohomology with $h^{2,1}_+ = 1$ and $h^{2,1}_- = 0$. This even/odd splitting of hodge numbers ensures that there are three K\"ahler moduli $T_\alpha$ and two involutively odd axions $G^a$. Further, there will not be any complex structure moduli, however a vector multiplet will appear in the four dimensional ${\cal N}=1$ effective theory due to non-trivial $(2,1)$-even sector as $h^{2,1}=h^{2,1}_+=1$. 

Now, the three involutively even- and two odd-harmonic (1,1)-forms can be written in the following manner \cite{Robbins:2007yv}, 
\begin{eqnarray}
 && \mu_1 = \frac{i}{4}\, \left(dz^1 \wedge d \ov{z}^1 +dz^2 \wedge d \ov{z}^2 \right) = dx^1 \wedge dx^2 + dx^3 \wedge dx^4 \\
 && \mu_2 = \frac{i}{2\sqrt{2}}\, \left(dz^1 \wedge d \ov{z}^1 -dz^2 \wedge d \ov{z}^2 \right) = dx^1 \wedge dx^3+dx^1 \wedge dx^4 - dx^2 \wedge dx^3 + dx^2 \wedge dx^4\nonumber\\
 && \mu_3 = \frac{i}{2}\, \left(dz^3 \wedge d \ov{z}^3 \right) = dx^5 \wedge dx^6 \nonumber
\end{eqnarray}
and
\begin{eqnarray}
 && \nu_1 = \frac{1-i}{4}\, \left(dz^1 \wedge d \ov{z}^2 + i \, d\ov{z}^1 \wedge d {z}^2 \right) = dx^1 \wedge dx^3- dx^1 \wedge dx^4 + dx^2 \wedge dx^3 + dx^2 \wedge dx^4 \nonumber\\
 && \nu_2 = -\frac{e^{-i \pi /4}}{4}\, \left(dz^1 \wedge d \ov{z}^2 - i \, d\ov{z}^1 \wedge d {z}^2 \right) = dx^1 \wedge dx^2 - dx^3 \wedge dx^4.
\end{eqnarray}
The respective even/odd dual four-forms can be written as under,
\begin{eqnarray}
 & & \hskip-2cm {\rm Even:} \hskip1cm \tilde{\mu}^1 = \mu_1 \wedge \mu_3, \, \, \, \tilde{\mu}^2 = \mu_2 \wedge \mu_3, \, \, \, \tilde{\mu}^3 = \frac{1}{2} \, \mu_1 \wedge \mu_1 \nonumber\\
 & & \hskip-2cm {\rm Odd:} \hskip1cm \tilde{\nu}^1 = \nu_1 \wedge \mu_3, \, \, \, \tilde{\nu}^2 = \nu_2 \wedge \mu_3
\end{eqnarray}
The toroidal orientifold under consideration also has a single non-trivial six-form 
\begin{eqnarray}
 \Phi_6 = dx^1 \wedge dx^2 \wedge dx^3 \wedge dx^4 \wedge dx^5 \wedge dx^6
\end{eqnarray}
while there is no harmonic 1-form and the dual five-form. For the present setup, the details of various non-vanishing intersection numbers defined in eqn. (\ref{eq:intersection}), are given as under \cite{Robbins:2007yv}
\bea
\label{eq:intersectionForm}
& & f=\frac{1}{4}, \, \, \hat{d}_\alpha^\beta = diag\left(\frac{1}{2},-1,\frac{1}{4}\right), \, \, \, {d}_a^b = diag\left(-1,-\frac{1}{2}\right)\nonumber\\
& & \left(k_{113}=\frac{1}{2}, \, k_{223}=-1\right)  \, \, \, \, \, {\rm and} \, \, \, \, \, \, \left(\hat{k}_{311}=-1, \, \hat{k}_{322}= -\frac{1}{2}\right).
\eea
Now, as one can expand the (1,1)-K\"ahler form $J$ as $J= t^1 \, \mu_1 +t^2 \, \mu_2 +t^3 \, \mu_3$ from eqn. (\ref{eq:fieldExpansions}), therefore using the intersection numbers given in eqn. (\ref{eq:intersectionForm}), the volume of the sixfold in Einstein frame is simplified as,
\begin{eqnarray}
\label{eq:volumeString}
 {\cal V}_E \equiv \frac{1}{3 !} \int_{X} J \wedge J \wedge J\, = \frac{1}{4} \, t^3 \, \left({(t^1)}^2 - 2 {(t^2)}^2\right) 
\end{eqnarray}
where the K\"ahler cone conditions for Einstein frame two-cycle volume moduli are given as $t^1 >0, \, t^3 > 0 , \, (t^1)^2 > 2 (t^2)^2$ to ensure the positive definiteness of the overall volume. 

\section{Scalar potential and search of new generalized flux orbits}
\label{sec_potential}
The four-dimensional scalar potential receives contributions from F-terms and D-terms, which we discuss in detail now. Subsequently, we will come to the search of some new generalized flux orbits at the end of this section. 
\subsection{F-term contributions}The F-term contributions to the ${\cal N}=1$ scalar potential are computed from the K\"ahler and super-potential via
\bea
\label{VF}
V_{F}=e^{K}\Big(K^{i\bar\jmath}D_i W\, D_{\bar\jmath} \ov W-3
|W|^2\Big)\,.
\eea
\subsubsection*{Writing the K\"ahler potential ($K$)}
To express the K\"ahler potential in terms of chiral variables, we have to rewrite the volume expression (\ref{eq:volumeString}). Note that, the last term in $T_\alpha$ represents the Einstein frame valued volume of the even four-cycles, and can be expressed in terms of the two-cycle volumes $t^\alpha$'s. For that purpose, a simplified version of chiral variables $T_\alpha$ is,
\bea
T_\alpha=-i \, \left(\frac{1}{2}\, \kappa_{\alpha\beta\gamma}\, t^\beta t^\gamma -\frac{1}{2} e^{-\phi} \, \kappa_{\alpha a b}\, b^a b^b \right)+  \, \left({\rho}_\alpha +\hat{\kappa}_{\alpha a b} \, {c^a b^b} + \frac{1}{2}\, \, \, C_0 \hat{\kappa}_{\alpha a b} \, {b^a b^b}\right) \, ,
\eea
which using $C_0 = c_0$, $e^{-\phi} = s$ and intersection numbers given in eqn. (\ref{eq:intersectionForm}) results in 
\bea
& & T_1 = -i \,  \, t_1 \, t_3 + \, \rho_1, \, \, \, \, T_2 = -i \,  \, t_2 \, t_3 + \, \rho_2 \\
& & T_3 = -i \,  \, \biggl[ \left(t_1^2 - 2 \, t_2^2\right) + s \, (2 b_1^2 + b_2^2) \biggr] + \biggl(\rho_3 - 4 b_1 \, c_1 - 2 b_2 \,c_2 -c_0 \,(2 \, b_1^2 + b_2^2) \biggr). \nonumber
\eea
{\it From now onwards we switch the upper indices in $t^\alpha$'s and $b^a/c^a$'s to the lower places for simplicity in presentation}. Considering $Im(T_\alpha) = -\tau_\alpha$ results in
\bea
& & \hskip-1.5cm  \tau_1= t_1 \, t_3 := \sigma_1, \, \tau_2 =  t_2 \,  t_3 := \sigma_2, \,\\
& &  \tau_3 =\left(t_1^2 - 2 \, t_2^2\right) + s \, (2 \, b_1^2 + b_2^2) :=\sigma_3 + s \, (2 \, b_1^2 + b_2^2), \nonumber
\eea
where we have also expressed Einstein-frame quantities $\sigma_\alpha := \frac{1}{2}\, \kappa_{\alpha\beta\gamma}\, t^\beta t^\gamma$ in terms of $\tau_\alpha$'s. Subsequently, the overall volume given in eqn. (\ref{eq:volumeString}) can be rewritten as below,
\begin{eqnarray}
 {\cal V}_E =  \frac{1}{4} \, \, \, \sqrt{\tau_1^2 -2 \, \tau_2^2}\, \sqrt{\tau_3 - 2 \, s\, b_1^2 - s \, b_2^2} \equiv \frac{1}{4} \sqrt{\sigma_1^2 - 2 \sigma_2^2} \, \sqrt{\sigma_3}
\end{eqnarray}
Now, the Einstein frame internal metric is
\bea
\label{eq:metric6DE}
     g_{ij}^E= \left(
\begin{array}{cccccc}
 {t^1} & 0 & {t^2} & -{t^2} & 0 & 0 \\
 0 & {t^1} & {t^2} & {t^2} & 0 & 0 \\
 {t^2} & {t^2} & {t^1} & 0 & 0 & 0 \\
 -{t^2} & {t^2} & 0 & {t^1} & 0 & 0 \\
 0 & 0 & 0 & 0 & {t^3} & 0 \\
 0 & 0 & 0 & 0 & 0 & {t^3} \\
\end{array}
\right)
\eea
which can be rewritten in terms of $\tau_\alpha$'s by using the relations: $t_1 = \frac{4 \, {\cal V}_E \, \tau_1}{\tau_1^2-2\, \tau_2^2}, \, \, t_2 = \frac{4 \, {\cal V}_E \, \tau_2}{\tau_1^2-2\, \tau_2^2}$ and $t_3 = \frac{4 \, {\cal V}_E}{\tau_3- s \, (2 \, b_1^2 + b_2^2)}$. Note that, the NS-NS axions appear in the internal metric while the same being written in terms of $\tau_\alpha$'s. Further, these four-cycle volumes $\tau_\alpha$'s have to be further expressed in terms of appropriate ${\cal N} =1$ coordinates $\{\tau, T_\alpha, G^a\}$ given as follows,
\bea
\label{eq:Volume}
& & {\cal V}_E\equiv {\cal V}_E(T_\alpha, S, G^a) = \frac{1}{4}\, \, \left(\frac{i(T_3 -\ov T_3)}{2} - \frac{i}{4 (\tau -\ov \tau)} \, \hat{\kappa}_{3 a b} \, (G^a -\ov G^a) (G^b -\ov G^b)\right)^{1/2} \nonumber\\
& & \hskip2cm \times  \biggl[\biggl(\frac{i(T_1 -\ov T_1)}{2}\biggr)^2 -2\, \biggl(\frac{i(T_2 -\ov T_2)}{2}\biggr)^2\biggr]^{1/2} \, \, 
\eea
Given that $h^{2,1}_-(X) =0$ in the present case, the complex structure moduli dependent part of the tree level K\"ahler potential defined in (\ref{eq:K}) is just a constant piece which can be nullified via an appropriate normalization $\left(i\int_{X}\Omega_3\wedge{\bar\Omega_3}\right)=1$. For example, we can consider ${\cal Z}^0 = 1$ and ${\cal F}_0=-i$, and subsequently the canonically normalized holomorphic three-form $\Omega_3$ given in (\ref{eq:Omega3main}) can be expressed as,
\bea
\label{eq:Omega3simp1}
\Omega_3&= \frac{1}{\sqrt 2}\, \left({\cal A}_0 + i \, {\cal B}^0\right)\, .
\eea 
An appropriate normalization is important to make, and will be crucial later on while establishing the match among the two scalar potentials; one computed from $K$ and $W$ (plus D-terms) while the other one coming from the dimensional reduction of a 10D oxidized conjectural form. Now, by using the volume form (\ref{eq:Volume}), the simplified K\"ahler potential expression to be heavily utilized later simplifies down to the form,
\bea
\label{typeIIBK}
K &= -\ln\left(-i(\tau-\ov \tau)\right) - 2 \, \ln{\cal V}_E(T_\alpha, \tau, G^a; \ov T_\alpha, \ov \tau, \ov G^a)
\eea

\subsubsection*{Writing the superpotential ($W$)}
Using eqn. (\ref{eq:Omega3simp1}) for canonically normalized holomorphic three-form $\Omega_3$, the generic non-geometric flux superpotential expression given in (\ref{eq:Wmain}) simplifies as under,
\bea
\label{eq:Wsimp1}
& & \hskip-1.75cm W = -\frac{1}{\sqrt2}\biggl[\left(f_0 + \tau\, h_0 + \omega_{a0} \, G^a + \hat{Q}^{\alpha}{}_{0} \, T_\alpha\right) -i\,  \left(f^0 + \tau\, h^0 + \omega_a{}^0 \, G^a + \hat{Q}^{\alpha 0} \, T_\alpha\right)\biggr]\,,
\eea
where indices are summed with $\alpha = 1,2,3$ and $a = 1,2$ corresponding to three even (complexified) divisor volume moduli and two odd-axions. Now, one can compute the F-term scalar potential using this superpotential (\ref{eq:Wsimp1}) and the K\"ahler potential given in (\ref{typeIIBK}). Note that, although when considered in real six dimensional basis, there are 10 independent geometric flux ($\omega_{ij}{}^k$) as well as 10 independent non-geometric flux ($Q^{ij}{}_k$) components which are allowed by the orientifold projection as detailed in appendix \ref{sec_components}, however for fluxes counted by the complex indices, this superpotential (\ref{eq:Wsimp1}) effectively involves only 4 geometric flux ($\omega_a^0, \omega_{a 0}$) components and 6 non-geometric flux components $(\hat{Q}^{\alpha 0}, \hat{Q}^\alpha{}_0)$. In fact as we will see later, there are additional 6 geometric flux components ($\omega_\alpha{}^1, \omega_{\alpha 1}$) and 4 non-geometric flux components $(Q_a{}^1,Q_{a 1})$ with complex-index which appear via D-term. Here one should recall that $k = 0, K = 1, a = 1,2$ and $\alpha= 1,2,3$.

\subsection{D-term contributions}
In the presence of a non-trivial sector of even (2,1)-cohomology, i.e. for $h^{2,1}_+(X)\ne 0$, there are additional D-term contributions to the four dimensional scalar potential. Following the strategy of \cite{Robbins:2007yv}, the same can be determined via considering the following gauge transformations of RR potentials $C_{RR} = C_0 + C_2 + C_4$,
\bea
\label{eq:gauge1}
& & \hskip-1.0cm C_{RR} = \left(c_0 + c^a \nu_a + \rho_\alpha \tilde{\mu}^\alpha + U^K \wedge a_K + U_K \wedge b^K + D^\alpha_2\wedge \mu_\alpha\right) \\
& & \longrightarrow  C_{RR} + {\cal D} (\lambda^K a_K + \lambda_K b^K )\nonumber
\eea
Recall that the pair $(U_K, U^K)$ appear in the expansion of RR four-form $C_4$ as given in Eqn. (\ref{eq:fieldExpansions}). The dimensional reduction of RR four-form on three-cycles can induce the relevant gauge fields in the four dimensional theory. Now using the flux actions on symplectic basis $(a_K, b^K)$, the second line of eqn. (\ref{eq:gauge1}) can be expanded as under,
\bea
\label{eq:gauge2}
& & \hskip-0.50cm C_{RR} + {\cal D} (\lambda^K a_K + \lambda_K b^K ) \\
& & \hskip-0.5cm = D^\alpha_2\wedge \mu_\alpha + \left(c_0 - f^{-1} R_K \lambda^K + f^{-1} R^K \lambda_K\right)  + \left(c^b -(d^{-1})_a{}^b Q^a{}_K \lambda^K + (d^{-1})_a{}^b Q^{a K} \lambda_K\right) \nu_b \nonumber\\
& & \hskip-0.50cm + \left(\rho_\alpha -({\hat{d}}^{-1})_\alpha{}^\beta \, \hat{\omega}_{\beta K} \lambda^K + ({\hat{d}}^{-1})_\alpha{}^\beta \, \hat{\omega}_{\beta}{}^{K} \lambda_K\right) \tilde{\mu}^\alpha + \left((U^K + d \lambda^K) \wedge a_K + (U_K + d \lambda_K)\wedge b^K\right)\nonumber
\eea
Note that the pair $(\lambda_K,\lambda^K)$ also ensures the 4D gauge transformations of quantities $(U_K, U^K)$ as $U^K \rightarrow U^K + d \lambda^K$ and $U_K \rightarrow U_K + d \lambda_K$.  Recollection of various pieces as given in eqn. (\ref{eq:gauge2}) implies a shift in the respective RR axionic parts of the chiral variables $\{\tau, G^a, T_\alpha\}$ via a redefinition of $c_0, c^a$ and $\rho_\alpha$ respectively. Subsequently the relevant variations of the chiral variables $\tau, G^a$ and $T_\alpha$ are given as,
\bea
\label{eq:gauge3}
& & \hskip1cm \delta \tau \equiv \delta c_0 = - f^{-1} R_K \lambda^K + f^{-1} R^K \lambda_K, \quad \\
& & \hskip1cm \delta G^a \equiv \delta c^a = -(d^{-1})_a{}^b Q^a{}_K \lambda^K + (d^{-1})_a{}^b Q^{a K} \lambda_K \nonumber\\
& & \hskip1cm \delta T_\alpha  \equiv \delta \rho_\alpha = -({\hat{d}}^{-1})_\alpha{}^\beta \, \hat{\omega}_{\beta K} \lambda^K + ({\hat{d}}^{-1})_\alpha{}^\beta \, \hat{\omega}_{\beta}{}^{K} \lambda_K \nonumber
\eea
Following the strategy of \cite{Binetruy:2004hh, Choi:2005ge}, and given that the superpotential (\ref{eq:W1}) is neutral under the gauge transformation (\ref{eq:gauge1}), the D-terms can be computed through the K\"ahler derivatives and variation of chiral fields (\ref{eq:gauge3}) via $D_i = i \, (\partial_A K) (\delta \phi^A_i)$ where $\phi_A = \{\tau, G^a, T_\alpha\}$ and $\delta \phi^A = \lambda^i(\delta \phi^A_i) + \lambda_i(\delta \phi^{Ai})$. This results in the following D-terms,
\bea
& & \hskip-1cm D_K = -i \, \biggl[ f^{-1} R_K \, (\partial_\tau K) + (d^{-1})_b{}^a Q^b{}_K \, (\partial_a K) + ({\hat{d}}^{-1})_\alpha{}^\beta \, \hat{\omega}_{\beta K}\, (\partial^\alpha K) \biggr] \\
& & \hskip-1cm D^K = i \, \biggl[ f^{-1} R^K \, (\partial_\tau K) + (d^{-1})_b{}^a Q^{b K} \, (\partial_a K) + ({\hat{d}}^{-1})_\alpha{}^\beta \, \hat{\omega}_{\beta}{}^{K}\, (\partial^\alpha K) \biggr] \nonumber
\eea
Note that we have both types of D-terms ($D_K, D^K$) unlike \cite{Robbins:2007yv} as we have not performed the symplectic transformations to rotate away half of the D-terms, namely $D^K$. These two D-term pieces  contribute to the four dimensional scalar potential in the following manner \cite{Robbins:2007yv}, 
\bea
& & {V_D}^{(1)} = \frac{1}{2} (Re \, \, \, {{\mathfrak G}})^{{-1}{JK}} \, D_J D_K + \frac{1}{2} (Re \, \, \, {\tilde{\mathfrak G}})^{-1}{}_{{JK}} \, D^J D^K \, ,
\eea
where $(Re \, \, \, {{\mathfrak G}})^{{-1}{JK}}$ and $(Re \, \, \, {\tilde{\mathfrak G}})^{-1}{}_{{JK}}$ represents the electric and magnetic gauge-kinetic couplings. These can be determined by considering the holomorphic three-form before orientifolding, say $\Omega_3^{(0)}$ which can be given as,
\bea
\label{eq:OmegaBefore}
& & \Omega_3^{(0)} = {\cal Z}^k \, {\cal A}_k - {\cal F}_k \, {\cal B}^k +  {\cal X}^K \, {a}_K - {\cal G}_K \, b^K
\eea
where ${\cal F}_k$ and ${\cal G}_K$ are both considered to be functions of ${\cal Z}^k$ and ${\cal X}^K$ arising from ${\cal N}=2$ prepotential before orientifolding is done. The electric gauge kinetic coupling is given by \cite{Grimm:2004uq},
\bea
\label{eq:gaugecouplings}
& & {\mathfrak G}_{KJ} = -\frac{i}{2} \, \left(\frac{\partial}{\partial {\cal X}^K} \, {{\cal G}_J}\right)_{at \, \, {\cal X}^K = 0}
\eea
Similarly, magnetic gauge kinetic couplings, $\tilde{\mathfrak G}$ are computed by interchanging $a_K$ and $b^K$ by a symplectic transformation. Note that, gauge kinetic couplings (${\mathfrak G}$ and $\tilde{\mathfrak G}$) are holomorphic functions of complex structure moduli. Now using the expressions for the generic tree level K\"ahler potential (\ref{typeIIBK}), one finds that \cite{Grimm:2004uq}
\bea
& & \hskip-1.75cm \partial_\tau K = \frac{i}{2\, s\, {\cal V}_E} \left({\cal V}_E - \frac{s}{2}\hat{k}_{\alpha a b} t^\alpha b^a b^b \right), \quad \\
& & \hskip-2.0cm \partial_{G^a} K = \frac{i}{2 \, {\cal V}_E}\hat{k}_{\alpha a b} t^\alpha b^b, \quad \partial_{T_\alpha} K = -\frac{i \, {\hat{d}}^\alpha{}_\beta \, t^\beta}{2 \, {\cal V}_E} \nonumber
\eea
Subsequently, we have
\bea
\label{eq:DtermOld}
& & \hskip-1.2cm \quad D_K = \frac{1}{2\, s\,{\cal V}_E}\, \biggl[ \frac{R_K}{f} \, \left({\cal V}_E -\frac{s}{2}\hat{k}_{\alpha a b} t^\alpha b^a b^b\right) + s\, (d^{-1})_b{}^a Q^b{}_K \, \hat{k}_{\alpha a c} t^\alpha b^c - s\, t^\alpha \, \,\hat{\omega}_{\alpha K}\, \biggr]\\
& & \hskip-1.2cm \quad D^K = -\frac{1}{2\, s\,{\cal V}_E}\, \biggl[ \frac{R^K}{f} \, \left({\cal V}_E -\frac{s}{2}\hat{k}_{\alpha a b} t^\alpha b^a b^b\right) + s\, (d^{-1})_b{}^a Q^{b K} \, \hat{k}_{\alpha a c} t^\alpha b^c - s\, t^\alpha \, \,\hat{\omega}_{\alpha}{}^{K}\, \biggr]\nonumber
\eea
This form of $D$-term suggests the use of some new flux combinations as we will discuss later.

\subsection{Intuitive search for the generalized flux orbits}
Let us perform an intuitive search for the correct flux combinations in the form of {\it new generalized flux orbits} modified by the presence of odd axions $B_2$ and $C_2$. Later on, we will show how our conjectured form of the new flux orbits is useful for a rearrangement of the total scalar potential via explicit calculation. For that purpose, we look into the superpotential components via the following three-form factor
\bea
& & \left({F} +\tau \, {H} + \, \omega_a \, {G}^a + \, \hat{Q}^{\alpha} \, \,{T}_\alpha \right)\nonumber\\
& & \quad \quad =\quad \left(F^k + \tau\, H^k + \omega_a{}^k \, G^a + \hat{Q}^{\alpha k} \, T_\alpha\right){\cal A}_k +\left(F_k + \tau\, H_k + \omega_{ak} \, G^a + \hat{Q}^{\alpha}{}_{k} \, T_\alpha\right){\cal B}^k \nonumber
\eea
Now using the expansion of chiral variables we can club the different pieces into the following manner,
\bea
& &  \hskip-1cm \left({F} +\tau \, {H} + \, \omega_a \, {G}^a + \, \hat{Q}^{\alpha} \, \,{T}_\alpha \right) \\
& & \quad =\biggl[{\mathbb F}^k + i \, \left(s \, {\mathbb H}^k \right)- i\, \left(\hat{\mathbb{Q}}^{\alpha k}\, \, \sigma_\alpha \right) \biggr] \, {\cal A}_k \, +\biggl[{\mathbb F}_k + i \, \left(s \, {\mathbb H}_k \right)- i\, \left(\hat{\mathbb Q}^{\alpha}{}_{k}\, \,\sigma_\alpha \right) \biggr] \, {\cal B}^k \, , \nonumber
\eea
where the symbol $\sigma_\alpha$ represents Einstein-frame four cycle volume given as: $\sigma_\alpha = \frac{1}{2}\,{\kappa}_{\alpha \beta \gamma} t^\beta t^\gamma$, and we propose the following flux combinations which generalize the Type IIB orientifold results of \cite{Blumenhagen:2013hva} with the inclusion of odd axions, 
\bea
\label{eq:orbitsA1}
& & {\mathbb H}_k \equiv {h}_k, \quad \quad \quad \hat{\mathbb{Q}}^{\alpha}{}_{k}\, = \hat{Q}^{\alpha}{}_{k}, \quad \quad \quad {\mathbb F}_k \equiv {f}_k +c_0\, {h}_k \\
& & {\mathbb H}^k \equiv {h}^k, \quad \quad \quad \hat{\mathbb{Q}}^{\alpha k}\, = \hat{Q}^{\alpha k}, \quad \quad \quad {\mathbb F}^k \equiv {f}^k +c_0\, {h}^k \nonumber
\eea
where
\bea
\label{eq:orbitsA2}
& &  \hskip-0.9cm {h}_k = H_k + \omega_{ak} \, {b}^a +\hat{Q}^\alpha{}_k \, \left(\frac{1}{2}\, \hat{\kappa}_{\alpha a b} b^a b^b\right), \nonumber\\
& & \hskip2cm \quad  {h}^k = H^k + \omega_{a}{}^{k} \, {b}^a + \hat{Q}^{\alpha k} \, \left(\frac{1}{2}\, \hat{\kappa}_{\alpha a b} b^a b^b\right),  \nonumber\\
& &  \hskip-0.9cm {f}_k = F_k + \omega_{ak} \, {c}^a + \hat{Q}^\alpha{}_k \, \left({\rho}_\alpha + \hat{\kappa}_{\alpha a b} c^a b^b\right) , \\
& & \hskip2cm \quad  {f}^k = F^k + \omega_a{}^k \, {c}^a + \hat{Q}^{\alpha k} \, \left({\rho}_\alpha + \hat{\kappa}_{\alpha a b} c^a b^b\right)\,. \, \nonumber
\eea
This is interesting to observe that similar to type IIA compactification on ${\mathbb T}^6/({\mathbb Z}_2 \times {\mathbb Z}_2)$-orientifold case \cite{Blumenhagen:2013hva}, the $H_3$ flux is receiving corrections of $(\omega \triangleleft B_2)$- and $\hat{Q}\triangleright (B_2 \wedge B_2)$-type, also in the type IIB orientifold case. However, the same will not have a correction of $R\bullet (B_2 \wedge B_2 \wedge B_2)$-type because, such terms will involve intersection numbers $\hat{\kappa}_{abc}$ which are zero by orientifold construction itself. Also, while invoking the new flux orbits, we find that RR flux, $F_3$ is having a correction of $(\omega \triangleleft C_2)$- as well as $\hat{Q}\triangleright (C_4 + C_2 \wedge B_2)$-type. 

Now, motivated by the type IIA generalized flux orbits proposed in \cite{Blumenhagen:2013hva}, it is tempting to guess that odd-indexed geometric flux components ($\omega_{ak},\omega_{a}{}^{k}$) will receive contributions of type $Q\triangleright B_2$ as under, 
\bea
\label{eq:orbitsA3}
& & {\mathbb\mho}_{ak} = \omega_{ak} + \hat{Q}^\alpha{}_k \, \left(({\hat{d}^{-1}})_\alpha^{ \, \,\delta} \, \hat{k}_{\delta a b}\, b^b\right) \, \quad {\mathbb\mho}_{a}{}^{k} = \omega_{a}{}^{k} + \hat{Q}^{\alpha k} \, \left(({\hat{d}^{-1}})_\alpha^{ \, \,\delta} \, \hat{k}_{\delta a b}\, b^b\right) \nonumber\\
& & \hskip3cm \hat{\mathbb Q}^\alpha{}_k = \hat{Q}^\alpha{}_k , \quad \hat{\mathbb Q}^{\alpha k} = \hat{Q}^{\alpha k} \,,
\eea
However orientifold invariance does not allow for the presence of non-geometric R-fluxes in new geometric flux components ${\mathbb\mho}_{ak}$ and ${\mathbb\mho}_{a}{}^{k}$.

Now let us also see if there is a possibility of combining other fluxes to construct corrections for geometric-flux orbits with even-indexed ($K\in h^{2,1}_+(X)$) components. For that, we observe that we can rewrite the D-terms in eqn. (\ref{eq:DtermOld}) relevant for $V_{D}^{(1)}$ in the following manner,
\bea
\label{eq:D-termCompact}
& &  D_K  = \frac{1}{2\, s\,{\cal V}_E}\, \biggl[ f^{-1} R_K \, {\cal V}_E - s\, t^\alpha \, \hat{\mho}_{\alpha K} \biggr]\, , \\
& &  D^K  = -\frac{1}{2\, s\,{\cal V}_E}\, \biggl[ f^{-1} R^K \, {\cal V}_E - s\, t^\alpha \,\hat{\mho}_{\alpha}{}^{K} \biggr] \, , \nonumber
\eea
where the generalized version of geometric flux components are collected as under,
\bea
& & \hat{{\mathbb\mho}}_{\alpha K} = \hat{\omega}_{\alpha K}\, - (d^{-1})_b{}^a  \, Q^{b}{}_{K} \, \, \, \left(\hat{k}_{\alpha a c} \, b^c \right) + f^{-1} \, \, R_K \, \left(\frac{1}{2}\hat{k}_{\alpha a b} \, b^a \,b^b\right)\\
& & \hat{{\mathbb\mho}}_{\alpha}{}^{K} =\hat{\omega}_{\alpha}{}^{K}\, - (d^{-1})_b{}^a \, \, Q^{b K} \, \, \, \left(\hat{k}_{\alpha a c} \, b^c\right)+ f^{-1} \, \, R^K \, \left(\frac{1}{2}\hat{k}_{\alpha a b} \, b^a \,b^b\right)\nonumber
\eea
Therefore, we have a generalized version of the even/odd components of geometric flux, and for non-geometric flux it can be analogously given as under,
\bea
& & \hskip-2cm  \hat{\mho}_\alpha \equiv (\mho \triangleleft \mu_\alpha) = \hat{\mho}_\alpha{}^K a_K + \hat{\mho}_{\alpha{}K} b^K, \, \, \, \, \,  {\mathbb Q}^{a}\equiv ({\mathbb Q} \triangleright \tilde{\nu}^a) = {\mathbb Q}^{a{}K} \, a_K + {\mathbb Q}^{a}{}_{K} b^K
\eea
where 
\bea
& & {\mathbb Q}^{a{}K} = {Q}^{a{}K} - f^{-1} \,\, d_b{}^a  \, (R^K  \, b^b)\, , \, \, \,  {\mathbb Q}^{a}{}_{K} =  {Q}^{a}{}_{K} - f^{-1} \,\,d_b{}^a \, (R_K \,b^b).
\eea
In \cite{Shukla:2015rua}, a modular completion of all these NS-NS and RR flux orbits have been proposed with the inclusion of P-fluxes which are S-dual to non-geometric Q-fluxes.

\section{Suitable rearrangement of scalar potential and dimensional oxidation}
\label{sec_oxidation}
Now, we will represent the four dimensional scalar potential into suitable pieces by utilizing our new generalized flux orbits and subsequently we will look for the possibility of oxidizing those pieces into ten dimensions. Here we will rewrite the full scalar potential in a particular form. The reasons for this rearrangement are as followings,
\begin{itemize}
\item{The well known Bianchi identities expressed with background fluxes written in real six dimensional indices are given as \cite{Shelton:2005cf},
\bea
\label{eq:bianchids1}
        {H}_{m[\underline{ab}} {\omega}^{m}{}_{\underline{cd}]}&=0  \nonumber\\
         {\omega}^{m}{}_{[\underline{bc}}  \, {\omega}^{d}{}_{\underline{a}]m}- {H}_{m[\underline{ab}} \, {Q}_{\underline{c}]}{}^{md}&=0\nonumber\\
      {\omega}^{m}{}_{[\underline{ab}]} \, {Q}_{m}{}^{[\underline{cd}]} -
     4\, {\omega}^{[\underline{c}}{}_{m[\underline{a}} \, {Q}_{\underline{b}]}{}^{\underline{d}]m} + {H}_{mab} \, {R}^{mcd}  &=0\, \\
       {Q}_{m}{}^{[\underline{bc}}  \, {Q}_{d}{}^{\underline{a}]m}- {R}^{m[\underline{ab}} \, \,
  {\omega}^{\underline{c}]}{}_{md}&=0\nonumber\\
      {R}^{m[\underline{ab}} \, {Q}_{m}{}^{\underline{cd}]}&=0  , \, \nonumber
\eea 
where underlined indices are anti-symmetrized. Now, one has to compute the total scalar potential by converting all fluxes, appearing in the superpotential eqn. (\ref{eq:Wsimp1}) and D-term eqn. (\ref{eq:D-termCompact}),  into real index components such as ($H_{ijk}, \omega_{ij}{}^k, Q^{ij}{}_k, R^{ijk} $ and ${F_{ijk}}$). Subsequently, we can use this set of Bianchi identities (\ref{eq:bianchids1}) to simplify the total potential. 
}
\item{The subsequent representation of scalar potential is what we call a 'suitable' rearrangement, as it will be directly useful for invoking its ten-dimensional origin.}
\end{itemize}
Fortunately, for the current toroidal setup, we can convert the superpotential (\ref{eq:Wsimp1}) as well as the D-term (\ref{eq:D-termCompact}) expressions into the ones written with real indexed flux components. This is the beauty of simplicity of toroidal models in which one can analytically compute all the relevant data including the internal six dimensional metric (unlike a generic CY case) for performing an explicit computation.
\subsubsection*{Rewriting the new generalized flux orbits}
Let us first recall the various flux orbits and summarize those at one place. The flux orbits in NS-NS sector with orientifold odd-indices $k \in h^{2,1}_-(X)$ are given as,
\bea
\label{eq:OddOrbitA}
& &  {\mathbb H}_k = H_k + \omega_{ak} \, {b}^a + \hat{Q}^\alpha{}_k \, \left(\frac{1}{2}\, ({\hat{d}^{-1}})_\alpha^{ \, \,\delta} \, \hat{k}_{\delta a b}\, b^a b^b\right) \nonumber\\
& &  {\mathbb H}^k = H^k + \omega_{a}{}^{k} \, {b}^a + \hat{Q}^{\alpha k} \, \left(\frac{1}{2}\, ({\hat{d}^{-1}})_\alpha^{ \, \,\delta} \, \hat{k}_{\delta a b}\, b^a b^b\right)\\
& & \hskip-2cm {\mathbb\mho}_{ak} = \omega_{ak} + \hat{Q}^\alpha{}_k \, \left(({\hat{d}^{-1}})_\alpha^{ \, \,\delta} \, \hat{k}_{\delta a b}\, b^b\right), \quad {\mathbb\mho}_{a}{}^{k} = \omega_{a}{}^{k} + \hat{Q}^{\alpha k} \, \left(({\hat{d}^{-1}})_\alpha^{ \, \,\delta} \, \hat{k}_{\delta a b}\, b^b\right) \nonumber\\
& & \hskip2cm \hat{\mathbb Q}^\alpha{}_k = \hat{Q}^\alpha{}_k , \quad \hat{\mathbb Q}^{\alpha k} = \hat{Q}^{\alpha k} \nonumber
\eea
while the flux components of even-index $K\in h^{2,1}_+(X)$ are given as, 
\bea
\label{eq:OddOrbitB}
& & \hat{\mho}_{\alpha K} = \hat{\omega}_{\alpha K}\, - (d^{-1})_b{}^a  \, Q^{b}{}_{K} \, \, \left(\hat{k}_{\alpha a c} \, b^c \right) + f^{-1} \, \, R_K \, \left(\frac{1}{2}\hat{k}_{\alpha a b} \, b^a \,b^b\right)\nonumber\\
& & \hat{\mho}_{\alpha}{}^{K} =\hat{\omega}_{\alpha}{}^{K}\, - (d^{-1})_b{}^a \, \, Q^{b K} \, \left(\hat{k}_{\alpha a c} \, b^c\right)+ f^{-1} \, \, R^K \, \left(\frac{1}{2}\hat{k}_{\alpha a b} \, b^a \,b^b\right)\nonumber\\
& & \hskip-1.5cm {\mathbb Q}^{a}{}_{K} =  {Q}^{a}{}_{K} - f^{-1} \,\,d_b{}^a \, (R_K\, b^b), \quad {\mathbb Q}^{a{}K} = {Q}^{a{}K} - f^{-1} \,\, d_b{}^a  \, (R^K \, \, b^b), \\
& & \hskip3cm {\mathbb R}_K = R_K, \quad {\mathbb R}^K = R^K \, .\nonumber
\eea
The RR three-form flux orbits are generalized in the following form, 
\bea
& &  \hskip-0.9cm {f}_k = F_k + \omega_{ak} \, {c}^a + \hat{Q}^\alpha{}_k \, \left({\rho}_\alpha + \hat{\kappa}_{\alpha a b} c^a b^b\right) , \\
& & \hskip2cm \quad  {f}^k = F^k + \omega_a{}^k \, {c}^a + \hat{Q}^{\alpha k} \, \left({\rho}_\alpha + \hat{\kappa}_{\alpha a b} c^a b^b\right)\,. \, \nonumber
\eea
Let us also mention that the action of various geometric as well as non-geometric fluxes on a given $p$-form, $X_p = \frac{1}{p!} X_{i_1 ....i_p} dx^1 \wedge dx^2 ....\wedge dx^{p}$, can be equivalently defined as under \cite{Robbins:2007yv},
\bea
& & (\omega \triangleleft X)_{i_1i_2...i_{p+1}} = \left(\begin{array}{c}p+1\\2\end{array}\right) \, \, \omega_{[\underline{i_1 i_2}}{}^{j} X_{j|\underline{i_3.....i_{p+1}}]} + \frac{1}{2} \left(\begin{array}{c}p+1\\1\end{array}\right) \, \, \omega_{j[\underline{i_1}}{}^{j} X_{\underline{i_2 i_3.....i_{p+1}}]}\nonumber\\
& & (Q \triangleright X)_{i_1i_2...i_{p-1}} = \frac{1}{2}\left(\begin{array}{c}p-1\\1\end{array}\right) \, \, Q^{jk}{}_{[\underline{i_1}} X_{jk|\underline{i_2.....i_{p-1}}]} + \frac{1}{2} \left(\begin{array}{c}p-1\\0\end{array}\right) \, \, Q^{jk}{}_{j} X_{k |i_1 i_2.....i_{p+1}}\nonumber\\
& & (R \bullet X)_{i_1i_2...i_{p-3}} = \frac{1}{3!}\left(\begin{array}{c}p-3\\0\end{array}\right) \, \, R^{jkl} X_{jkl|i_1.....i_{p-3}]} \, ,
\eea 
where underlined indices are anti-symmetrized. Moreover, one can notice that the action of (non-)geometric-fluxes via $\triangleleft$, $\triangleright$ and $\bullet$ on a $p$-from changes the same into a $(p+1)$-form, a $(p-1)$-form and a $(p-3)$-form respectively. Using these generic definitions, the three-forms pieces, $({\omega}_a\,G^a)$ and $(Q^{\alpha} \, \, \, \,{T}_\alpha)$ appearing in the superpotential (\ref{eq:W1}) are expanded as under, 
\bea
& & \hskip-0.9cm (\omega_a  \,\, {G}^a)={1\over 3!} (\omega_a \,\, {G}^a)_{ijk}\, dx^i\wedge dx^j\wedge dx^k, \\
& & \quad ( {\hat{Q}}^{\alpha} \, \, \,  \,{T}_\alpha)={1\over 3!} (\hat{Q}^{\alpha} \, \, \, \,{T}_\alpha)_{ijk}\,dx^i\wedge dx^j\wedge dx^k \nonumber
\eea
where
\bea
\label{eq:genactions}
& & (\omega_a \, \, \,  {G}^a)_{ijk}={3}\, \, {\omega}_{[\underline{i k}}{}^{m\,}
    G_{m\underline{k}]}\,+{3\over 2}\, \, {\omega}_{{m [\underline i}}{}^{m\,}
    G_{\underline{jk}]}\, \\
& & (\hat{Q}^{\alpha} \,\,\, \,{T}_\alpha)_{ijk}={3\over 2}\, {Q}_{[\underline{i}}{}^{mn\,}
    T_{mn\underline{jk}]}\,+ {1\over 2}\, {Q}_{{m}}{}^{mn\,}
    T_{n[\underline{ijk}]}\,. \nonumber
\eea
The details of the enumeration of various flux and moduli/axions's components are summarized in the appendix \ref{sec_components}, and guided by the type II orientifold results of \cite{Blumenhagen:2013hva}, one finds that the even/odd-indexed flux components can be equivalently combined as follows,
\bea
\label{eq:orbitfluxes}
& & {\mathbb H}_{ijk}={H}_{ijk}+{3}\, \, {\omega}_{[\underline{i k}}{}^{m\,} B_{m\underline{k}]}\, -3\, {Q}_{[\underline{i}}{}^{mn} B_{m\underline{j}}\, B_{n\underline{k}]} \nonumber\\
& & {\mathbb \mho}^i{}_{jk}={\omega}^i{}_{jk}-2\,{Q}_{[\underline{j}}{}^{mi}  B_{m\underline{k}]}\,  -{R}^{mni}  B_{m[\underline{j}} B_{n\underline{k}]}\nonumber\\
& & {\mathbb Q}_k{}^{ij}={Q}_k{}^{ij} - {R}^{ijk'}\, B_{k' k}  , \hskip 4cm \{i, j\} \in \{1, 2,.., 6\}\nonumber\\
& & {\mathbb R}^{ijk}={R}^{ijk}\, . 
\eea
Here we also point out that, these flux orbits are very similar to those of type IIA compactified on ${\mathbb T}^6/({{\mathbb Z}_2\times{\mathbb Z}_2})$ orientifold \cite{Blumenhagen:2013hva} except an additional pieces $\ov{R}^{mnp}  B_{m[\underline{i}} B_{n\underline{j}} B_{p\underline{k}]}$ contributing to the ${\mathbb H}$-flux orbit. One should note again that $R^{lmn} B_{l[i} B_{mj} B_{nk]}$ piece of ${\mathbb H}$-flux orbit trivially vanishes as a reflection of the fact that intersection number $\hat{k}_{abc}$ with all three indices being odd, vanishes by the orientifold construction itself.  Further, despite of the presence of flux components of kind ${\omega}_{{m i}}{}^{m\,}$ and ${Q}_{{m}}{}^{mn\,}$, in present setup, we find that contributions of type ${\omega}_{{m [\underline i}}{}^{m\,} B_{\underline{jk}]}$ as well as ${Q}_{{m}}{}^{mn\,} B_{n[\underline{i}} \, B_{\underline{jk}]}$ to the flux orbits, which could have been expected from the most generic definitions in (\ref{eq:genactions}), are simply zero. 

\subsubsection*{Rewriting the superpotential ($W$)}
In our present setup, the overall structure gets much simpler because of the absence of complex structure moduli as $h^{2,1}_-(X) = 0$. This helps in writing both of the symplectic cohomology bases $({\cal A}_k, {\cal B}^k)$ and $({a}_K, {b}^K)$  as a constant linear combination of elements of real cohomology basis $({\alpha}_I, {\beta}^J)$ given as under \cite{Robbins:2007yv}
\bea
\label{eq:symplecticBases}
& & a_1 = -\frac{i}{2} \left(dz^1\wedge dz^2\wedge d {\ov z}^3-d {\ov z}^1\wedge d {\ov z}^2 \wedge d {z}^3 \right) = \beta^0 + \beta^1 + \beta^2 - \beta^3 \nonumber\\
& & b^1 = \frac{1}{2} \left(dz^1\wedge dz^2\wedge d {\ov z}^3+d {\ov z}^1\wedge d {\ov z}^2 \wedge d {z}^3 \right) = \alpha_0 + \alpha_1 + \alpha_2 - \alpha_3 \\
& & A_0 = \frac{1}{2} \left(dz^1\wedge dz^2\wedge d {z}^3+d {\ov z}^1\wedge d {\ov z}^2 \wedge d {\ov z}^3 \right) = \alpha_0 - \alpha_1 - \alpha_2 - \alpha_3 \nonumber\\
& & B^0 = -\frac{i}{2} \left(dz^1\wedge dz^2\wedge d {z}^3-d {\ov z}^1\wedge d {\ov z}^2 \wedge d {\ov z}^3 \right) = -\beta^0 + \beta^1 + \beta^2  + \beta^3 \nonumber
\eea
where the following notation have been considered,
\bea
\label{eq:Realbasis}
       \alpha_0&=dx^1\wedge dx^3\wedge dx^5\,, \qquad\qquad
       \beta^0=dx^2\wedge dx^4\wedge dx^6\, ,\nonumber\\
       \alpha_1&=dx^1\wedge dx^4\wedge dx^6\, , \qquad\qquad
       \beta^1=dx^2\wedge dx^3\wedge dx^5\, ,\\
       \alpha_2&=dx^2\wedge dx^3\wedge dx^6\, , \qquad\qquad
       \beta^2=dx^1\wedge dx^4\wedge dx^5\, ,\nonumber\\
      \alpha_3&=dx^2\wedge dx^4\wedge dx^5\, ,\qquad\qquad
      \beta^3=dx^1\wedge dx^3\wedge dx^6\,. \nonumber
\eea
Subsequently, one can represent all the NS-NS flux components as $H_{ijk}, \omega_{ij}{}^k, Q^{ij}{}_k, R^{ijk} $ and RR flux components as ${F_{ijk}}$. In this new basis we have,
\bea
\label{eq:Omega3simp}
\Omega_3&= \frac{1}{\sqrt{2}} \,\biggl[\left(\alpha_0 - i \, \beta^0\right)- \left(\alpha_1 - i \, \beta^1\right)-\left(\alpha_2 - i \, \beta^2\right)-\left(\alpha_3 - i \, \beta^3\right)\biggr] ,
\eea
where $\int \alpha_I\wedge \beta^J= - f\, \delta_I{}^J$ following from the definition of integration over the six-form $\Phi_6$ given in eqn. (\ref{eq:intersection}). The normalization $i\int_X \Omega_3 \wedge {\ov \Omega}_3 = 1$ remains intact as $f =1/4$ for the present orientifold. 
After utilizing the various non-vanishing components of all the (non-)geometric fluxes, the explicit form of superpotential (\ref{eq:Wsimp1}) becomes
\bea
\label{typeIIBWsimp1}
& &  \hskip-1cm W = \sqrt{2} \times \biggl[\biggl(F_{246} + \tau H_{246} + G^2 \left(-\omega_{15}{}^1 + \omega_{16}{}^1 + \omega_{25}{}^1 + \omega_{26}{}^1\right) + G^1 (-\omega_{35}{}^1 + \omega_{46}{}^1 )\nonumber\\
& & \quad \quad + (Q^{15}{}_4 + Q^{16}{}_3) \, T_1 + (Q^{15}{}_1 - Q^{15}{}_2 - Q^{16}{}_1 - Q^{16}{}_2)\, T_2 - Q^{13}{}_6 \, T_3\biggr)\\
& & \quad \quad + \, i \, \biggl(F_{135}  + \tau\, H_{135} - G^2 (\omega_{15}{}^1 +\omega_{16}{}^1 + \omega_{25}{}^1 - \omega_{26}{}^1) - G^1 (\omega_{36}{}^1 + \omega_{45}{}^1) \nonumber\\
& & \quad \quad - (Q^{15}{}_3 - Q^{16}{}_4)\, T_1 + (Q^{15}{}_1+ Q^{15}{}_2+ Q^{16}{}_1- Q^{16}{}_2)\, T_2 + Q^{13}{}_5\, T_3\biggr) \biggr]\,.\nonumber
\eea
Now, with the expansion known, it is easy to make the following connections for the two superpotential expressions (\ref{eq:Wsimp1}) and (\ref{typeIIBWsimp1}) which are the same \cite{Robbins:2007yv}, 
\bea
& & \hskip-1cm \omega_{a0} \equiv \left(\begin{array}{c}\omega_{15}{}^1 - \omega_{16}{}^1 -\omega_{25}{}^1 - \omega_{26}{}^1\\ \omega_{35}{}^1 -\omega_{46}{}^1\end{array}\right), \, \omega_{a}{}^{0} \equiv \left(\begin{array}{c}-\omega_{15}{}^1 -\omega_{16}{}^1 - \omega_{25}{}^1 + \omega_{26}{}^1\\ -\omega_{36}{}^1 - \omega_{45}{}^1\end{array}\right) \nonumber
\eea
and
\bea
& & \hskip-1.3cm \hat{Q}^{\alpha}{}_{0} \equiv \left(\begin{array}{c} -Q^{15}{}_4 - Q^{16}{}_3 \\ -Q^{15}{}_1 + Q^{15}{}_2 + Q^{16}{}_1 + Q^{16}{}_2 \\ Q^{13}{}_6\end{array}\right), \, \hat{Q}^{\alpha 0} \equiv \left(\begin{array}{c} -Q^{15}{}_3 + Q^{16}{}_4 \\ Q^{15}{}_1+ Q^{15}{}_2+ Q^{16}{}_1- Q^{16}{}_2 \\ Q^{13}{}_5\end{array}\right)\, .\nonumber
\eea
\subsubsection*{Rewriting the D-term scalar potential $V_D^{(1)}$}
For computing the D-term contribution to the scalar potential, we first need to know the holomorphic gauge kinetic couplings. For that let us follow the strategy of  \cite{Robbins:2007yv} by considering the expansion of holomorphic three-form $\Omega_3$ before the orientifold projection has been made. In this case, the single complex structure modulus appears as a deformation in one of the coordinates of the complex threefold via $z^3 = x^5 + U \, x^6$. Subsequently, using the definitions of $z^1$ and $z^2$ from eqn. (\ref{eq:coordinates}) along with the modified $z^3$ coordinated as above, we find that,
\bea
& & \hskip-1.5cm dz^1 \wedge dz^2 \wedge dz^3 = \biggl[\left( \alpha_0 + i \, U\, \alpha_1  + i \, U\, \alpha_2 -\alpha_3\right) + \left(-U \, \beta^0 + i\, \beta^1 + i \, \beta^2 + U\, \beta ^3\right) \biggr]\, ,
\eea
where we have used the definitions of $\alpha_i$ and $\beta^j$ as given in eqn. (\ref{eq:Realbasis}). Further, using eqn. (\ref{eq:symplecticBases}), we can rewrite the above form in terms of the complex bases of even/odd (2,1)-cohomology as,
\bea
& & dz^1 \wedge dz^2 \wedge dz^3 = \frac{1-i\, U}{2} \biggl[ \left({\cal A}_0 + i\, {\cal B}^0\right) + \frac{i - U}{1- i\, U} (a_1 - i\, b^1)\biggr]\, .
\eea
Under the orientifold projection, the complex structure modulus gets fixed as $U = i$, and therefore the second half piece corresponding to the even (2,1)-cohomology bases vanishes. Recalling the fact that we have fixed the normalization after the orientifold projection in such a way that $\Omega_3^{(-)} = \frac{1}{\sqrt 2} \, \left({\cal A}_0 + i \, {\cal B}^0\right)$, and for having a consistent normalization throughout, we can trace back the appropriate expression of the holomorphic three-form $\Omega_3$ in the present case as under,
\bea
& &\hskip-1.5cm  \Omega_3^{(0)} =  \frac{\sqrt 2}{1-i\, U} \, dz^1 \wedge dz^2 \wedge dz^3 = \frac{1}{\sqrt 2} \biggl[ {\cal A}_0 + i\, {\cal B}^0 + \frac{i - U}{1- i\, U} (a_1 - i\, b^1) \biggr]. 
\eea
Now comparing the above form with the generic one as given in the eqn. (\ref{eq:OmegaBefore}) we find that ${\cal G}_1 = i\, {\cal X}^1$, and after using eqn. (\ref{eq:gaugecouplings}), we get
\bea
& & {\mathfrak G}_{11} = -\frac{i}{2} \, \left(\frac{\partial}{\partial {\cal X}^1} \, {{\cal G}_1}\right)_{at \, \, {\cal X}^1 = 0} = \frac{1}{2}\, .
\eea
Subsequently, using the expressions (\ref{eq:orbitfluxes}) of flux orbits and the constant gauge kinetic coupling being $1/2$, one gets the following additional pieces in the total $D$-term contributions \cite{Robbins:2007yv}, 
\bea
& & \hskip-0.5cm V_{D}^{(1)} = \frac{1}{s^2 \, {\cal V}_E^2} \biggl[\left(4\, {\cal V}_E + \, t_3 \, \, (2 \,s\, b_1^2 +s\, b_2^2) \right)\, R^{246} + t_3 \, s\, b_1 \, (Q^{15}{}_1-Q^{15}{}_2+Q^{16}{}_1+Q^{16}{}_2)\nonumber\\
& &   \quad + t_3 \, s \, b_2 \, (Q^{15}{}_3 + Q^{16}{}_4) - t_1 \, s \, (\omega_{36}{}^1-\omega_{45}{}^1) -t_2 \, s \, (-\omega_{15}{}^1-\omega_{16}{}^1+\omega_{25}{}^1-\omega_{26}{}^1)  \nonumber\\
& & \quad -t_3 \, s\, \omega_{14}{}^5\biggr]^2 + \frac{1}{s^2 \, {\cal V}_E^2} \biggl[\left(4\, {\cal V}_E + \, t_3 \, \, (2 \,s \, b_1^2 + s\,b_2^2) \right)\, R^{135} - t_3 \, s \, b_2 \, (Q^{15}{}_4 - Q^{16}{}_3) \nonumber\\
& & \quad  - t_3 \, b_1 \, s\, (-Q^{15}{}_1-Q^{15}{}_2+Q^{16}{}_1-Q^{16}{}_2) - t_1 \, s \,(\omega_{35}{}^1+\omega_{46}{}^1) \nonumber\\
& & \quad -t_2 \, s\, (-\omega_{15}{}^1+\omega_{16}{}^1-\omega_{25}{}^1-\omega_{26}{}^1) -t_3 \, s\, \omega_{13}{}^5\biggr]^2 \, ,\nonumber
\eea
From this, one has following relations of the even-indexed flux components in the matrix formulation \cite{Robbins:2007yv},
\bea
& & \hskip-0.5cm \hat{\omega}_{\alpha}{}^{1} \equiv \left(\begin{array}{c}\omega_{35}{}^1 +\omega_{46}{}^1\\ -\omega_{15}{}^1 + \omega_{16}{}^1 -\omega_{25}{}^1 - \omega_{26}{}^1\\ -\omega_{13}{}^5\end{array}\right), \, \hat{\omega}_{\alpha 1} \equiv \left(\begin{array}{c}\omega_{36}{}^1 -\omega_{45}{}^1\\-\omega_{15}{}^1 -\omega_{16}{}^1 + \omega_{25}{}^1 - \omega_{26}{}^1\\ \omega_{14}{}^5\end{array}\right) \nonumber
\eea
and
\bea
& & \hskip-1.0cm {Q}^{a1} \equiv \left(\begin{array}{c} -Q^{15}{}_1 - Q^{15}{}_2 + Q^{16}{}_1 - Q^{16}{}_2 \\ -Q^{15}{}_4 + Q^{16}{}_3\end{array}\right), \, {Q}^{a}{}_1 \equiv \left(\begin{array}{c} Q^{15}{}_1- Q^{15}{}_2+ Q^{16}{}_1+ Q^{16}{}_2 \\ Q^{15}{}_3 + Q^{16}{}_4\end{array}\right) \,.\nonumber
\eea
\subsection{Rewriting the four dimensional scalar potential}
Now, using these flux orbits (\ref{eq:orbitfluxes}), let us write the following pieces, which we will verify to be a `suitable' rearrangement of the total scalar potential subject to satisfying a set of Bianchi identities (\ref{eq:bianchids1}),
\bea
\label{eq:detailedV1}
& & \hskip-1.0cm {V_{{\mathbb H}{\mathbb H}}}= \frac{s}{{\cal V}_E} \, \biggl[\frac{1}{3!}\, {\mathbb H}_{ijk}\,  {\mathbb H}_{i'j'k'}\, g_E^{ii'}\, g_E^{jj'} g_E^{kk'}\biggr]\nonumber\\
& & \hskip-1.0cm {V_{{\mathbb Q}{\mathbb Q}}}= \frac{1}{s \, {\cal V}_E} \, \biggl[3 \times \left(\frac{1}{3!}\, {\mathbb Q}_k{}^{ij}\, {\mathbb Q}_{k'}{}^{i'j'}\,g^E_{ii'} g^E_{jj'} g_E^{kk'} \right) + \, 2 \times \left(\frac{1}{2!}{\mathbb Q}_m{}^{ni}\, {\mathbb Q}_{n}{}^{mi'}\, g^E_{ii'}\right) \biggr]\nonumber\\
& & \hskip-1.0cm {V_{{\mathbb H}{\mathbb Q}}}= \frac{1}{{\cal V}_E} \, \biggl[{\bf (+2)} \times \left(\frac{1}{2!} {\mathbb H}_{mni} \, {\mathbb Q}_{i'}{}^{mn}\, g_E^{ii'}\right)\biggr] \\
& & \hskip-1.0cm {V_{{\mathbb F}{\mathbb F}}}= \frac{1}{{\cal V}_E}\biggl[\frac{1}{3!}\, {\mathbb F}_{ijk}\,  {\mathbb F}_{i'j'k'}\, g_E^{ii'} \, g_E^{jj'} g_E^{kk'}\biggr] \nonumber\\
& & \hskip-1.0cm {V_{{\mathbb H}{\mathbb F}}}=  \frac{1}{{\cal V}_E} \biggl[{\bf (+2)} \times \left(\frac{1}{3!} \, \times\, \frac{1}{3!} \, \, {\mathbb F}_{ijk} \,\, {\cal E}_E^{ijklmn} \, \, {\mathbb H}_{lmn}\right) \biggr] \equiv {\rm Generalized \, \, tadpoles}\nonumber\\
& & \hskip-1.0cm {V_{{\mathbb F}{\mathbb Q}}}= \frac{1}{s \, {\cal V}_E} \, \, \Big[{\bf (+2)} \times \biggl( \frac{1}{4!} \, \times\, \frac{1}{2!} \, {\mathbb Q}_{i}{}^{j'k'} \, {\mathbb F}_{j'k'j} \, \, \, \, \sigma^E_{klmn} \,\,\, {\cal E}_E^{ijklmn} \, \,  \biggr)\Big] \equiv {\rm Generalized \, \, tadpoles}\nonumber
\eea
and
\bea
\label{eq:detailedV2}
& & \hskip-3.5cm {V_{{\mathbb R}{\mathbb R}}}= \frac{1}{s^2\, {\cal V}_E}\biggl[\frac{1}{3!}\, {\mathbb R}^{ijk}\, {\mathbb R}^{i'j'k'}\, g^E_{ii'} \, g^E_{jj'} g^E_{kk'}\biggr] \\
& & \hskip-3.5cm {V_{{\mathbb \mho}{\mathbb \mho}}}= \frac{1}{{\cal V}_E} \, \biggl[3 \times \left(\frac{1}{3!}\, {\mathbb \mho}_{ij}{}^k\, {\mathbb \mho}_{i'j'}{}^{k'}\,g_E^{ii'} g_E^{jj'} g^E_{kk'} \right) + \, 2 \times \left(\frac{1}{2!}{\mathbb \mho}_{ni}{}^m\, {\mathbb \mho}_{mi'}{}^n\, g_E^{ii'}\right) \biggr]\nonumber\\
& & \hskip-3.5cm {V_{{\mathbb R}{\mathbb \mho}}}= \frac{1}{s\, {\cal V}_E} \, \biggl[{\bf (+2)} \times \left(\frac{1}{2!} {\mathbb R}^{mni} \, \, \, {\mathbb \mho}_{mn}{}^{i'}\, g^E_{ii'}\right)\biggr] \nonumber
\eea
where ${\mathbb F}_{ijk}= \left(F_{ijk} +{3}\, \, {\omega}_{[\underline{i j}}{}^{m\,} C_{m\underline{k}]}\, -3\, {Q}_{[\underline{i}}{}^{mn} B_{m\underline{j}}\, C_{n\underline{k}]} + {3\over 2}\, {Q}_{[\underline{i}}{}^{mn\,}
    C^{(4)}_{mn\underline{jk}]} \right) + \, c_0 \, {\mathbb H}_{ijk}$ has been utilized.

In order to understand and appreciate the nice structures within the aforementioned expressions, we need to supplement the followings,
\begin{itemize}
\item{We have utilized some Einstein- and string-frame conversion relations given as ${\cal V}_E = s^{3/2}\, {\cal V}_s, \, g^E_{ij} =  g_{ij} \, \sqrt{s}$ and $g_E^{ij} = g^{ij}/\sqrt{s}$. The metric is given in eqn. (\ref{eq:metric6DE}).}
\item{The Levi-civita tensors are defined in terms of antisymmetric Levi-civita symbols $\epsilon_{ijklmn}$ and the same are given  as: ${\cal E}^E_{ijklmn}=\sqrt{|g_{ij}|} \, \,\epsilon_{ijklmn} = (4\, {\cal V}_E)\, \, \epsilon_{ijklmn}$ while ${\cal E}_E^{ijklmn}=\epsilon^{ijklmn}/\sqrt{|g_{ij}|} =\, \epsilon^{ijklmn}/(4\, {\cal V}_E) $. The presence of extra factor of $4$ is attributed to the intersection numbers in eqns. (\ref{eq:intersection})-(\ref{eq:intersectionForm}), and one has to take care of this throughout for dimensional oxidation process.
}
\item{Further, the symbol $\sigma^E_{klmn}$ denotes the Einstein-frame volume of the four-cycles written in components of the real 6D basis of the internal manifold.}
\end{itemize}
Now, we verify the claim that eqns. (\ref{eq:detailedV1}) and (\ref{eq:detailedV2}) indeed represent the same 4D scalar potential by providing intermediate connections. The first six pieces given in eqn. (\ref{eq:detailedV1}) consist of terms which come mostly from the F-term contribution $V_F$, while the last three pieces in eqn. (\ref{eq:detailedV2}) consist of terms which are mostly coming from (a part) of D-term contributions which was earlier mentioned as $V_D^{(1)}$. However, it is important to state that there is still some small mixing between these two sectors of $F$- and $D$-term contributions. 

The expressions of K\"ahler potential (\ref{typeIIBK}) and the superpotential (\ref{typeIIBWsimp1}) allow one to compute the effective four-dimensional scalar potential which results in 1302 number of terms via the F-term contributions. It is important to mention that due to the complicated nature of this orientifold setup, unlike the case of ${\mathbb T}^6/({\mathbb Z}_2 \times {\mathbb Z}_2)$, we do not have a well separated rearrangement of pieces to catch inside ${V_{F}}$ and ${V_{D}^{(1)}}$ independently. Nevertheless, we still find that some pieces are nicely separable as followings,
\bea
\label{eq:counting}
& & \biggl\{{V_{{\mathbb H}{\mathbb H}}}, \quad  {V_{{\mathbb F}{\mathbb F}}} , \quad  {V_{{\mathbb H}{\mathbb F}}}, \quad {V_{{\mathbb F}{\mathbb Q}}} \biggr\}\subset {V_{F}},  \\
& & \hskip-1.5cm \#({V_{{\mathbb H}{\mathbb H}}})= 76 \, , \quad \#({V_{{\mathbb F}{\mathbb F}}})= 520 \, \, \quad \#({V_{{\mathbb H}{\mathbb F}}})= 200 \, , \quad \#({V_{{\mathbb F}{\mathbb Q}}})= 292. \nonumber
\eea
Singling out such cleanly separable terms in pieces of $(\ref{eq:counting})$ takes care of a huge number of terms, and so helps a lot in analyzing the remaining terms. The counting of these terms goes such that out of a total of 1302 terms of F-term contribution, we are able to rearrange 1088 terms in what we call {\it a cleanly separable suitable form} (for oxidation purpose). Thus we are only left with 214 terms of $V_F$, which are clubbed to form other flux-orbits after being added with ${V_{D}^{(1)}}$, and leaving behind some terms canceled by Bianchi identities. The type of terms which could be captured into the form of what we call `suitable' rearrangement are indeed in the form as under,
\bea
\label{eq:potential1}
& & \hskip-1.5cm {V_{F}}+{V_{D}^{(1)}}= {V_{{\mathbb H}{\mathbb H}}}+ {V_{{\mathbb F}{\mathbb F}}} + {V_{{\mathbb H}{\mathbb F}}}+ {V_{{\mathbb F}{\mathbb Q}}}+ {V_{{\mathbb R}{\mathbb R}}}  \\
& & \hskip+1.5cm +  {V_{{\mathbb Q}{\mathbb Q}}} + {V_{{\mathbb \mho}{\mathbb \mho}}} + {V_{{\mathbb H}{\mathbb Q}}} + {V_{{\mathbb R}{\mathbb \mho}}} + ....... \quad ,\nonumber
\eea
where dots denote a collection of terms which are canceled by using the Bianchi identities ({\ref{eq:bianchids1}}). Interestingly, we find that $R$-flux contributions coming from D-term ${V_{D}^{(1)}}$ can be written in a very similar fashion to those of other pieces. Note that, although the terms ${V_{{\mathbb H}{\mathbb Q}}}, {V_{{\mathbb \mho}{\mathbb \mho}}}, {V_{{\mathbb Q}{\mathbb Q}}}$ and ${V_{{\mathbb R}{\mathbb \mho}}}$ are not as cleanly separated, nevertheless they are indeed part of ${V_{F}}+{V_{D}^{(1)}}$ subject to satisfying a set of Bianchi identities ({\ref{eq:bianchids1}}). 

Following the strategy of \cite{Gao:2015nra}, we deliberately seek for topological terms ${V_{{\mathbb H}{\mathbb F}}}$ and ${V_{{\mathbb F}{\mathbb Q}}}$ in our rearrangement, because of the fact that such terms can be nullified via adding local source contributions such as brane/orientifold planes. Thus we propose additional D-term contributions for these local sources written with new generalized flux orbits to have a form as under,
\bea
& & {V_{D}^{(2)}} = - {V_{{\mathbb H}{\mathbb F}}} - {V_{{\mathbb F}{\mathbb Q}}} \supset \left\{{V_{{F}{H}}}, {V_{{F}{\omega}}}, {V_{{F}{Q}}}, {\rm BIs}\right\}
\eea
As it has been seen in \cite{Gao:2015nra} also, this piece ${V_{D}^{(2)}}$ not only has contributions from various 3/5/7-branes and 3/5/7-orientifolds but also involves some mixing of the other flux-squared pieces (killed via NS-NS Bianchi identities) while being written in terms of the new generalized flux orbits instead of usual generalized fluxes. 

Finally, we conclude this section with the following rearrangement of total four dimensional effective scalar potential subject to satisfying a (sub)set of Bianchi identities (\ref{eq:bianchids1}),
\bea
\label{eq:potential2}
& & \hskip-1.5cm V_{tot} \equiv {V_{F}}+{V_{D}^{(1)}} + {V_{D}^{(2)}} = {V_{{\mathbb H}{\mathbb H}}}+ {V_{{\mathbb F}{\mathbb F}}} + {V_{{\mathbb R}{\mathbb R}}}  +  {V_{{\mathbb Q}{\mathbb Q}}} + {V_{{\mathbb \mho}{\mathbb \mho}}} + {V_{{\mathbb H}{\mathbb Q}}} + {V_{{\mathbb R}{\mathbb \mho}}}, 
\eea
where various pieces are elaborated in eqns. (\ref{eq:detailedV1})-(\ref{eq:detailedV2}).

\subsection{Dimensional oxidation}
Following the strategy of \cite{Blumenhagen:2013hva, Gao:2015nra}, we are now in a position to propose a dimensional oxidation of the four dimensional scalar potential (\ref{eq:potential2}). The rearrangement of the total potential is already made to what we call a ``suitable'' form. Assuming all the fluxes to be constant parameters appearing as constant fluctuations around the internal background, now all we need to do is to fix the correct coefficients of the integral measure of the 10D kinetic terms. For that, we consider that the non-vanishing components of the 10D metric in string frame are
\bea
\label{eq:gMN}
    g_{MN}={\rm blockdiag}\Big({e^{2\phi}\over {\cal V}_s} \, \, \tilde g_{\mu\nu}, \, \, g_{ij}\Big) \, .
\eea
where $\tilde g_{\mu\nu}$ denote the 4D Einstein-frame metric. Subsequently, the ten-dimensional integral measure simplifies to,
\bea
& & \hskip-1.0cm \int  d^{10} x\,  \sqrt{-g} \, (....) \simeq \int  d^{4} x\, \sqrt{-g_{\mu\nu}} \left(\frac{1}{s^{4} \, {\cal V}_s^2}\right) \times \left(\int  d^{6} x\,  \sqrt{-g_{mn}}\right) \, \times (..........) ~~ \\
& & \hskip2.2cm \simeq \int  d^{4} x\, \sqrt{-g_{\mu\nu}} \times \left(\frac{4}{s^{4} \, {\cal V}_s}\right) \times (..........). \nonumber
\eea
as $\int  d^{6} x\,  \sqrt{-g_{mn}} \equiv 4\, {\cal V}_s$ gives the string-frame 6D volume by using the string-frame version of the metric components given in eqn. (\ref{eq:metric6DE}). Just to recall that a factor of $4$ appears due to choice of normalization following from the definition of integration over the six-form $\Phi_6$ given in eqn. (\ref{eq:intersection}) where $f =1/4$ in the current setup. Now the string frame version of the ten-dimensional action, which reproduces the four-dimensional scalar potential (\ref{eq:potential2}) upon a dimensional reduction, can be conjectured to have the following form,
\bea
\label{eq:oxiaction}
      & & \hskip-1.5cm  S={1\over 2}\int  d^{10} x\,  \sqrt{-g} \, \, \Big( {\bf {\cal L}_{\mathbb F \mathbb F}}+{\bf {\cal L}_{\mathbb H \mathbb H}}+{\bf {\cal L}_{\mathbb \mho \mathbb \mho}}  +{\bf {\cal L}_{\mathbb Q \mathbb Q}}+{\bf {\cal L}_{\mathbb R \mathbb R}} + {\bf {\cal L}_{{\mathbb H}{\mathbb Q}}} + {\bf {\cal L}_{{\mathbb R}{\mathbb \mho}}}  \Big)
\eea
where
\bea
\label{eq:detailedV0}
& & {\bf {\cal L}_{\mathbb H \mathbb H}}= -{e^{-2\phi}\over 2} \, \biggl[\frac{1}{3!}{\mathbb H}_{ijk}\,  {\mathbb H}_{i'j'k'}\, g^{ii'}\, g^{jj'} g^{kk'}\biggr]\, \nonumber\\
& & {\bf {\cal L}_{\mathbb \mho \mathbb \mho}}= -{e^{-2\phi}\over 2} \, \biggl[3 \times \left(\frac{1}{3!}\, {\mathbb \mho}_{ij}{}^k\, {\mathbb \mho}_{i'j'}{}^{k'}\,g^{ii'} g^{jj'} g_{kk'} \right)+ \, 2 \times \left(\frac{1}{2!}{\mathbb \mho}_{ni}{}^m\, {\mathbb \mho}_{mi'}{}^n\, g^{ii'}\right) \biggr]\nonumber\\
& & {\bf {\cal L}_{\mathbb Q \mathbb Q}}= -{e^{-2\phi}\over 2} \, \biggl[3 \times \left(\frac{1}{3!}\, {\mathbb Q}_k{}^{ij}\, {\mathbb Q}_{k'}{}^{i'j'}\,g_{ii'} g_{jj'} g^{kk'} \right)+ \, 2 \times \left(\frac{1}{2!}{\mathbb Q}_m{}^{ni}\, {\mathbb Q}_{n}{}^{mi'}\, g_{ii'}\right) \biggr]\nonumber\\
& & {\bf {\cal L}_{\mathbb R \mathbb R}}= -{{e^{-2\phi}}\over 2} \, \biggl[\frac{1}{3!}{\mathbb R}^{ijk}\, {\mathbb R}^{i'j'k'}\, g_{ii'} \, g_{jj'} g_{kk'}\biggr] \nonumber\\
& & {\bf {\cal L}_{{\mathbb H}{\mathbb Q}}}= -{e^{-2\phi}\over 2} \, \biggl[{\bf (+2)} \times \left(\frac{1}{2!} {\mathbb H}_{mni} \, {\mathbb Q}_{i'}{}^{mn}\, g^{ii'}\right)\biggr]\\
& & {\bf {\cal L}_{{\mathbb R}{\mathbb \mho}}}= -{e^{-2\phi}\over 2} \, \biggl[{\bf (+2)} \times \left(\frac{1}{2!} {\mathbb R}^{mni} \, {\mathbb \mho}_{mn}{}^{i'}\, g_{ii'}\right)\biggr]. \nonumber\\
& & {\bf {\cal L}_{\mathbb F \mathbb F}}= -{{1}\over 2} \, \biggl[\frac{1}{3!}{\mathbb F}_{ijk}\,  {\mathbb F}_{i'j'k'}\, g^{ii'} \, g^{jj'} g^{kk'}\biggr] \nonumber
\eea
Now, the (inverse-)metric components are written in string-frame. This completes our goal of implementing odd axions $B_2/C_2$ into the dimensional oxidation process proposed with non-geometric Q-fluxes in \cite{Blumenhagen:2013hva}, and further generalized with the dual P-fluxes in \cite{Gao:2015nra}. Moreover, the ten-dimensional pieces given in eqns. (\ref{eq:oxiaction}) and (\ref{eq:detailedV0}) can be further connected to the ten dimensional DFT action on the lines of \cite{Blumenhagen:2013hva}. 

\section{Conclusion}
\label{sec_conclusion}
Following the strategy of \cite{Blumenhagen:2013hva, Gao:2015nra},  we have implemented the presence of involutively odd-axions in the dimensional oxidation process. Considering an explicit example of type IIB compactification on an orientifold of ${\mathbb T}^6/{\mathbb Z}_4$ sixfold, we have first invoked a new version of generalized flux orbits previously proposed in \cite{Blumenhagen:2013hva} which have led to a possible rearrangement of the four dimensional scalar potential. This scalar potential has various (what we call) `suitable' pieces which suggest to conjecture a ten-dimensional non-geometric action. As opposed to the most of the previous studies with Type IIB compactification on ${\mathbb T}^6/({\mathbb Z}_2\times {\mathbb Z}_2)$-orientifold, this analysis with ${\mathbb T}^6/{\mathbb Z}_4$-orientifold has not only included odd-axions via having $h^{1,1}_-(X) \neq 0$ but at the same time, it has also incorporated the additional D-term contributions which helps in inclusion of non-geometric R-flux to have a broader framework having all NS-NS fluxes. This has been possible via considering the orientifold involution $\sigma$ such that $h^{2,1}_+(X) \neq 0$ as opposed to the standard approach of studying type IIB-orientifold compactification with $h^{2,1}(X)= h^{2,1}_-(X)$ in which cases, the $R$-fluxes could not be turned-on. In support of the proposal made in \cite{Blumenhagen:2013hva}, the ten dimensional pieces as given in eqns. (\ref{eq:oxiaction}) and (\ref{eq:detailedV0}) should be valid beyond the present toroidal model, and the dimensional reduction on a generic orientifold of a complex threefold should induce all the respective $F$- and $D$-term contributions (subject to satisfying a set of Bianchi identities) in the four dimensional scalar potential. On these lines,  this work may be considered as another step towards understanding the ten-dimensional origin of the most generic non-geometric 4D type IIB supergravity action equipped with all standard as well as (non-)geometric NS-NS and RR-fluxes, and we hope to get back to it in near future.

\section*{Acknowledgments}
I am very thankful to Ralph Blumenhagen for useful discussions and continuous encouragements. Moreover, I am thankful to Anamaria Font, Xin Gao, Daniela Herschmann, Oscar Loaiza-Brito and Erik Plauschinn for useful discussions. This work was supported by the Compagnia di San Paolo contract ``Modern Application of String Theory'' (MAST) TO-Call3-2012-0088.

\clearpage
\appendix

\section{Components of fluxes surviving under the orientifold involution}
\label{sec_components}
Here we recollect various components of fluxes and $p$-forms which survive under the orientifold involution \cite{Robbins:2007yv}, 
\begin{itemize}
\item{{\bf NS-NS $H_3$-flux:}
 \begin{eqnarray}
 & & H_{135}, \,  H_{245}, \,  H_{146}, \,  H_{236}, \,  H_{246}, \, H_{136}, \,  H_{145}, \, H_{235} \nonumber\\
 & & \hskip-2cm {\rm where} \nonumber\\
       & &  H_{135} = -  H_{245} = - H_{146}=- H_{236}, \\
       & &  H_{246} = -  H_{136} = - H_{145}=- H_{235} \nonumber
       \end{eqnarray}}
\item{{\bf R-R $F_3$-flux:}
 \begin{eqnarray}
 & &  F_{135}, \,  F_{245}, \,  F_{146}, \,  F_{236}, \,  F_{246}, \,  F_{136}, \,  F_{145}, \,  F_{235} \nonumber\\
 & & \hskip-2cm {\rm where} \nonumber\\
       & & F_{135} = -  F_{245} = - F_{146}=- F_{236}, \\
       & & F_{246} = -  F_{136} = - F_{145}=- F_{235} \nonumber
       \end{eqnarray}}
\item{{\bf Geometric $\omega_{ij}^k$-flux:}
 \bea
 & & \omega_{15}^{1}, \, \omega_{25}^{2}, \,\omega_{36}^{3}, \,\omega_{46}^{4}, \,\omega_{16}^{1}, \,\omega_{26}^{2}, \,\omega_{35}^{3}, \,\omega_{45}^{4}, \,\omega_{25}^{1}, \,\omega_{15}^{2}, \nonumber\\
 & & \,\omega_{46}^{3}, \,\omega_{36}^{4}, \,\omega_{26}^{1}, \,\omega_{16}^{2}, \,\omega_{45}^{3}, \,\omega_{35}^{4}, \,\omega_{35}^{1}, \,\omega_{45}^{2}, \,\omega_{26}^{3}, \,\omega_{16}^{4}, \nonumber\\
 & & \omega_{36}^{1}, \,\omega_{46}^{2}, \,\omega_{25}^{3}, \,\omega_{15}^{4}, \,\omega_{45}^{1}, \,\omega_{35}^{2}, \,\omega_{16}^{3}, \,\omega_{26}^{4}, \,\omega_{46}^{1}, \,\omega_{36}^{2}, \nonumber\\
 & & \omega_{15}^{3}, \,\omega_{25}^{4}, \,\omega_{13}^{5}, \,\omega_{24}^{5}, \,\omega_{14}^{6}, \,\omega_{23}^{6}, \,\omega_{14}^{5}, \,\omega_{23}^{5}, \,\omega_{13}^{6}, \,\omega_{24}^{6}
 \eea
 where
\bea
  & & \omega_{15}^{1}= - \, \omega_{25}^{2}=- \,\omega_{36}^{3}= \,\omega_{46}^{4}, \, \, \, \, \omega_{16}^{1}=- \,\omega_{26}^{2}= \,\omega_{35}^{3}=- \,\omega_{45}^{4}, \nonumber\\
  & & \omega_{25}^{1}= \,\omega_{15}^{2}=- \omega_{46}^{3}= - \,\omega_{36}^{4}, \, \, \, \,\omega_{26}^{1}= \,\omega_{16}^{2}= \,\omega_{45}^{3}= \,\omega_{35}^{4}, \nonumber\\
  & & \omega_{35}^{1}=- \,\omega_{45}^{2}=- \,\omega_{26}^{3}= -\,\omega_{16}^{4}, \, \, \, \,\omega_{36}^{1}=- \,\omega_{46}^{2}= \,\omega_{25}^{3}= \,\omega_{15}^{4}, \nonumber\\
  & & \omega_{45}^{1}= \,\omega_{35}^{2}= \,\omega_{16}^{3}=- \,\omega_{26}^{4}, \, \, \, \, \omega_{46}^{1}= \,\omega_{36}^{2}=- \omega_{15}^{3}= \,\omega_{25}^{4}, \nonumber\\
  & & \omega_{13}^{5}=- \,\omega_{24}^{5}= \,\omega_{14}^{6}= \,\omega_{23}^{6}, \, \, \, \,\omega_{14}^{5}= \,\omega_{23}^{5}=- \,\omega_{13}^{6}= \,\omega_{24}^{6} \nonumber
\eea }
\item{{\bf Non-geometric $Q^{ij}_k$-flux:}
 \bea
 & & Q^{15}_{1}, \, Q^{25}_{2}, \,Q^{36}_{3}, \,Q^{46}_{4}, \,Q^{16}_{1}, \,Q^{26}_{2}, \,Q^{35}_{3}, \,Q^{45}_{4}, \,Q^{25}_{1}, \,Q^{15}_{2}, \nonumber\\
 & & \,Q^{46}_{3}, \,Q^{36}_{4}, \,Q^{26}_{1}, \,Q^{16}_{2}, \,Q^{45}_{3}, \,Q^{35}_{4}, \,Q^{35}_{1}, \,Q^{45}_{2}, \,Q^{26}_{3}, \,Q^{16}_{4}, \nonumber\\
 & & Q^{36}_{1}, \,Q^{46}_{2}, \,Q^{25}_{3}, \,Q^{15}_{4}, \,Q^{45}_{1}, \,Q^{35}_{2}, \,Q^{16}_{3}, \,Q^{26}_{4}, \,Q^{46}_{1}, \,Q^{36}_{2}, \nonumber\\
 & & Q^{15}_{3}, \,Q^{25}_{4}, \,Q^{13}_{5}, \,Q^{24}_{5}, \,Q^{14}_{6}, \,Q^{23}_{6}, \,Q^{14}_{5}, \,Q^{23}_{5}, \,Q^{13}_{6}, \,Q^{24}_{6}
 \eea
 where
 \bea
        & & Q^{15}_{1}= - \, Q^{25}_{2}= \,Q^{36}_{3}=- \,Q^{46}_{4}, \, \, \, \, Q^{16}_{1}=- \,Q^{26}_{2}= -\,Q^{35}_{3}= \,Q^{45}_{4}, \nonumber\\
       & & Q^{25}_{1}= \,Q^{15}_{2}= Q^{46}_{3}=  \,Q^{36}_{4}, \, \, \, \,Q^{26}_{1}= \,Q^{16}_{2}= -\,Q^{45}_{3}= -\,Q^{35}_{4}, \nonumber\\
       & & Q^{35}_{1}= -\,Q^{45}_{2}= \,Q^{26}_{3}= \,Q^{16}_{4}, \, \, \, \,Q^{36}_{1}=- \,Q^{46}_{2}= - \,Q^{25}_{3}= -\,Q^{15}_{4}, \nonumber\\
       & & Q^{45}_{1}= \,Q^{35}_{2}= -\,Q^{16}_{3}= \,Q^{26}_{4}, \, \, \, \, Q^{46}_{1}= \,Q^{36}_{2}= Q^{15}_{3}= - \,Q^{25}_{4}, \nonumber\\
       & & Q^{13}_{5}=- \,Q^{24}_{5}=- \,Q^{14}_{6}=- \,Q^{23}_{6}, \, \, \, \,Q^{14}_{5}= \,Q^{23}_{5}= \,Q^{13}_{6}=- \,Q^{24}_{6} \nonumber
\eea}
\item{{\bf Non-geometric $R^{ijk}$-flux:}
 \begin{eqnarray}
 & &  R^{135}, \,  R^{245}, \,  R^{146}, \,  R^{236}, \,  R^{246}, \,  R^{136}, \,  R^{145}, \, R^{235} \nonumber\\
 & & \hskip-4cm {\rm where} \nonumber\\
       & &  R^{135} = -  R^{245} =  R^{146}=  R^{236}, \\
       & &  R^{246} = -  R^{136} =  R^{145}=  R^{235} \nonumber
       \end{eqnarray}}
\item{{\bf NS-NS $B_2$-field:}
 \begin{eqnarray}
 & &  B_{12}, \,  B_{13}, \, B_{14}, \, B_{23}, \, B_{24}, \, B_{34}, \, \nonumber\\
 & & \hskip-5cm {\rm where} \nonumber\\
       & &  B_{12} = -  B_{34} \equiv b_2, \\
       & &  B_{13} = -  B_{14} =  B_{23} =  B_{24} \equiv b_1 \nonumber
       \end{eqnarray}}
\item{{\bf R-R $C_2$-field:}
 \begin{eqnarray}
 & &  C_{12}, \,  C_{13}, \, C_{14}, \, C_{23}, \, C_{24}, \, C_{34}, \, \nonumber\\
 & & \hskip-5cm {\rm where} \nonumber\\
       & &  C_{12} = -  C_{34} \equiv c_2 , \\
       & &  C_{13} = -  C_{14} =  C_{23} =  C_{24} \equiv c_1 \nonumber
       \end{eqnarray}}
\item{{\bf R-R $C_4$-field:}
 \begin{eqnarray}
 & &  C_{1234}, \,  C_{1256}, \,  C_{3456}, \, C_{1356}, \, C_{2456}, \, C_{2356}, \, C_{1456}, \, \nonumber\\
 & & \hskip-3.5cm {\rm where} \nonumber\\
       & &  C_{1256} =  C_{3456} \equiv \rho_1, \\
       & &  C_{1356} =  C_{2456} =-  C_{2356} =  C_{1456} \equiv \rho_2 \nonumber\\
       & &  C_{1234} \equiv \rho_3 \nonumber
       \end{eqnarray}}
\end{itemize}

\newpage

\bibliographystyle{utphys}
\bibliography{reference}

\end{document}